\documentclass[11pt,letterpaper]{article}
\usepackage[left=1in,top=1in,right=1in,nohead]{geometry}
\usepackage{jheppub}
\usepackage{amsmath}
\usepackage{color}
\usepackage{graphicx}
\usepackage{hyperref}
\usepackage{slashed}
\usepackage{customcommands}
\usepackage[caption=false]{subfig}
\usepackage[justification=justified]{caption}

\begin{document}


\title{A Surprising Similarity Between Holographic CFTs and a Free Fermion in $(2+1)$ Dimensions}

\author{Krai Cheamsawat$^1$,}
\author{Sebastian Fischetti$^2$,}
\author{Lucas Wallis$^1$,}
\author{and Toby Wiseman$^1$}

\affiliation{$^1$Theoretical Physics Group, Blackett Laboratory, Imperial College, London SW7 2AZ, UK}
\affiliation{$^2$Department of Physics, McGill University, Montr\'eal, QC, H3A 2T8, Canada}

\emailAdd{krai.cheamsawat15@imperial.ac.uk}
\emailAdd{fischetti@physics.mcgill.ca}
\emailAdd{l.wallis17@imperial.ac.uk}
\emailAdd{t.wiseman@imperial.ac.uk}

\abstract{
We compare the behavior of the vacuum free energy (i.e.~the Casimir energy) of various~$(2+1)$-dimensional CFTs on an ultrastatic spacetime as a function of the spatial geometry.  The CFTs we consider are a free Dirac fermion, the conformally-coupled scalar, and a holographic CFT, and we take the spatial geometry to be an axisymmetric deformation of the round sphere.  The free energies of the fermion and of the scalar are computed numerically using heat kernel methods; the free energy of the holographic CFT is computed numerically from a static, asymptotically AdS dual geometry using a novel approach we introduce here.  We find that the free energy of the two free theories is qualitatively similar as a function of the sphere deformation, but we also find that the holographic CFT has a remarkable and mysterious \textit{quantitative} similarity to the free fermion; this agreement is especially surprising given that the holographic CFT is strongly-coupled.  Over the wide ranges of deformations for which we are able to perform the computations accurately, the scalar and fermion differ by up to 50\% whereas the holographic CFT differs from the fermion by less than one percent.
}

\maketitle

%
\section{Introduction}
\label{sec:intro}
%

Holography in the form of AdS/CFT~\cite{Maldacena:1997re,Witten:1998qj,Gubser:1998bc} is a very powerful tool for probing features of certain strongly-coupled field theories which would otherwise be intractable using conventional field-theoretic methods.  While in principle the dual gravitational description of such holographic field theories always exists as a full quantum theory of gravity, in practice holography can only give insights into the field theory when the gravitational theory is in a semiclassical regime.  This regime corresponds to a small Planck length in the bulk, and hence a large number of degrees of freedom in the field theory.  While the ``real'' field theories in which we might typically be interested do not have arbitrarily many degrees of freedom (nor necessarily even a holographic dual description), there has nevertheless been a large effort in using holography to describe physics for theories with relatively few degrees of freedom, or even in theories without holographic duals.  An example of such an application is to heavy ion physics motivated by the resemblance between QCD and certain Yang-Mills theories which do possess (at least conventional) dual gravitational descriptions, and under the assumption that three colors in QCD is ``close'' to large~$N$ in Yang-Mills for the purposes of computing observables of interest (see e.g.~\cite{CasalderreySolana:2011us} for a review).

When pushing holography beyond its obvious regime of validity, a phenomenological question then arises: how similar is the behavior of holographic field theories to conventional ones?  In this paper, we explore this question in the context of the computation of vacuum (i.e.~Casimir) energies of~$(2+1)$-dimensional CFTs.  Specifically, we compare the vacuum energy of two free CFTs -- the massless Dirac fermion and the conformally-coupled scalar -- to that of a holographic CFT.  These vacuum energies are non-local functionals of the geometry on which the CFTs live, and we are interested in their dependence on this geometry.  To that end, we take the CFTs to live on a product of time with a topological two-sphere.

While simple, this setting is of physical interest.  The free massless fermion is an effective description of the behavior of electrons in monolayer graphene (and generalizations thereof) near Dirac points, with the geometry on which the fermion lives given by the geometry of the monolayer~\cite{Graphene,GrapheneDirac1,GrapheneDirac2}.  The behavior of the vacuum energy of the free Dirac fermion on these geometries may therefore be relevant to technological applications.  From a different perspective, various statements about the Casimir energy and energy density in the holographic theory in this setting can be proved using elegant geometric arguments -- for example, for deformations of the two-sphere for which there exists an infilling bulk geometry, the Casimir energy is globally maximized by the round sphere~\cite{HicWis15}.  We have seen evidence this may also be true for free field theories~\cite{Fischetti:2020knf}, suggesting this might be a universal behavior among a large class of field theories, even beyond CFTs.

Now, for a CFT (as all the theories we consider in this paper are), the energy of the round sphere vanishes, and one is naturally led to ask how the energy varies as the sphere is deformed.  Since the dependence of the free energy on the area of the deformed sphere is trivial, we restrict to perturbations that leave this area unchanged.  For small such deformations, the Casimir energy is universal for all CFTs, depending only on their central charge~\cite{Bobev:2017asb,Fischetti:2017sut}\footnote{In the context of $(2+1)$-dimensional CFTs the constant we refer to as the ``central charge'' is the coefficient governing the two-point function of the stress tensor.}.  For large deformations, one would instead na\"ively expect different behaviors for different theories.  However, in~\cite{Fischetti:2020knf} it was observed that for fixed area deformations, the Casmir energies of the minimally-coupled scalar (which is \textit{not} a CFT) and of the Dirac fermion actually were not only always negative relative to the round sphere, but also qualitatively very similar.  Perhaps this is not too surprising, as the scalar and fermion are both free theories.  Our purpose in this paper is to compare the free field results to those of a holographic CFT, which on the other hand is strongly-coupled.

We will find that the behaviors of the Casimir energies of the scalar, fermion, and holographic CFTs are all qualitatively very similar.  In fact, our comparison of these energies yields a surprise: for a wide range of non-perturbative deformations of the sphere (with its area held fixed), the Casimir energy of the holographic theory is remarkably \textit{quantitatively} close to that of the free fermion.  While they are not identical, the surprising closeness of these energies is somewhat of a mystery and perhaps suggests the presence of some underlying mechanism responsible for this fine-tuning.  Moreover, such a close agreement between the fermion and the holographic CFT has also been exhibited in entropic computations like the entanglement entropy corner function~\cite{Bueno:2015rda,Bueno:2015xda} and the ratio~$c_s/c$, where~$c_s$ is the coefficient in the thermal entropy density of a CFT in flat space:~$s = c_s T^2$~\cite{Bueno:2015qya}.  We do note, however, that this similarity in the Casimir energy does not extend to other closely-related deformations.  For instance, if we were to turn on a finite temperature, the confining nature of the holographic CFT leads to a temperature-independent free energy at sufficiently low temperature with a corresponding first-order deconfinement transition~\cite{Witten:1998zw}, whereas the free theories will exhibit free energies with a nontrivial and smooth temperature dependence.  Likewise, in Euclidean signature the analogue of the free energy is the partition function, which was computed in~\cite{Bobev:2017asb} for these three CFTs on the Euclidean squashed (Berger) three-sphere (which doesn't have a Lorentzian interpretation).  In that case the fermion and holographic CFT yield different behaviors of the partition function, and in fact even the partition functions of the scalar and of the fermion are qualititatively different (for example, for large enough perturbations of the three-sphere, the partition function of the fermion is larger than that of the unsquashed sphere, while that of the scalar is always smaller).  These observations naturally prompt the question of why the fermion, scalar, and holographic CFT Casimir energies are all so similar for the classes of deformations we study in this paper.

A plan for the paper is as follows. First, in Section~\ref{sec:setting} we introduce the physical setup and the fixed-area sphere deformations we consider.  In Section~\ref{sec:gravity} we describe how we construct the bulk geometries dual to the vacuum state of the holographic CFT using the numerical methods of~\cite{Headrick:2009pv,Figueras:2011va,Wiseman:2011by}.  We also describe a novel method for extracting the vacuum energy much more accurately than the entire vacuum stress tensor itself.  In Section~\ref{sec:results} we then compare the Casimir energies of the various deformations we consider.  
%
%
We briefly conclude in Section~\ref{sec:disc}.  Appendix~\ref{app:convergence} contains details on the numerical approaches, including a brief review of the heat kernel computation of the vacuum energies of the free fields and an analysis of accuracy and convergence.

%
\section{Physical setting}
\label{sec:setting}
%

As remarked above, we take our~$(2+1)$-dimensional CFTs to be in vacuum on the spacetime
\be
\label{eq:metric}
ds^2 = - dt^2 + d\Sigma^2 = -dt^2 + h_{ij} dy^i \, dy^j,
\ee
where~$d\Sigma^2 = h_{ij} dy^i \, dy^j$ is the (time-independent) metric of a two-dimensional Riemannian geometry on a manifold with spherical topology; for simplicitly we will often denote this geometry as simply~$\Sigma$.  We do not turn on any other sources for local CFT operators.  Consequently, the vacuum energy~$E[\Sigma]$, also referred to as the Casimir energy, is a functional only of the geometry~$\Sigma$.  Note that because we are considering the vacuum state of CFTs, all dimensionful scales are set by the geometry; therefore a rigid rescaling~$d\Sigma^2 \to \lambda^2 d\Sigma^2$ (for some constant~$\lambda$) causes the vacuum energy to scale trivially as~$E[\Sigma] \to \lambda^{-1} E[\Sigma]$.  Hence without loss of generality we fix the area of~$\Sigma$ to be that of the unit round two-sphere:~$\mathrm{Area}[\Sigma] = 4 \pi$.

On the ultrastatic geometry~\eqref{eq:metric}, the vacuum energy~$E[\Sigma]$ can be defined as the integral over~$\Sigma$ of the energy density~$\langle T_{tt} \rangle$, where~$\langle T_{\mu\nu} \rangle$ is the vacuum expectation value of the stress tensor.  For a general QFT there are counterterm ambiguities when renormalizing the stress tensor on curved spacetime, but for a three-dimensional CFT the counterterms required by power counting -- corresponding to a cosmological constant and an Einstein-Hilbert term in the action -- both must vanish as they are not Weyl invariant.  Hence~$E[\Sigma]$ is an unambiguous physical quantity\footnote{
In fact, for a general QFT on the spacetime~\eqref{eq:metric}, one can consider the difference~$E[\Sigma] - E[S^2]$ between the vacuum energy of $\Sigma$ and that of the round two-sphere~$S^2$.  If~$\Sigma$ and~$S^2$ have the same area, this difference is finite and and removes all ambiguities due to counterterms
%
(for a diffeomorphism invariant UV regulator),
and is therefore physically meaningful~\cite{Fischetti:2018shp,Fischetti:2020knf}.
}.
We further note that the vacuum energy vanishes when~$\Sigma$ is the round sphere~$S^2$, which can be seen most easily by noting that the corresponding Euclidean spacetime~$\mathbb{R} \times S^2$ is conformally equivalent to Euclidean flat space~$\mathbb{R}^3$ on which the vacuum stress tensor vanishes.  The lack of any conformal anomaly means that the same must be true on the original geometry~$\mathbb{R} \times S^2$, and hence~$E[S^2] = 0$.

We will consider three CFTs: the conformally coupled scalar, the free Dirac fermion, and a holographic CFT in the regime in which its geometric dual is classical Einstein gravity (i.e.~the limit of large central charge and ``strong coupling'').  These have central charges~$c_s = 3/(4\pi)^2$,~$c_f = (3/2)/(4\pi)^2$, and~$c_h = \ell^2/(16\pi G)$, respectively, with~$\ell$ and~$G$ the dual AdS length and Newton's constant.  Because we are turning on no sources besides the metric, for the holographic CFT we will take the dual gravity theory to be four-dimensional pure gravity with negative cosmological constant, whose solutions may be embedded in various top-down models in explicit cases such as e.g.~the ABJM theory~\cite{ABJM}.  In principle there could be situations in which bulk matter fields spontaneously condense. However we will not consider such situations here.

Now, when~$\Sigma$ is a small deformation of the round sphere, we may write its metric in a form conformal to the round sphere as
\be
d\Sigma^2 =  \left(1 + 2 \epsilon f(\theta,\phi) \right) \left( d\theta^2 +  \sin^2{\theta} \, d\phi^2 \right),
\ee
with $| \epsilon | \ll 1$.  It is then possible to work perturbatively in~$\eps$ to derive the leading-order behavior of the vacuum energy: decomposing~$f$ in spherical harmonics as~$f = \sum_{l,m} f_{l,m} Y_{l,m}(\theta,\phi)$, one obtains~\cite{Fischetti:2017sut}
\be
\label{eq:Epert}
E[\Sigma] = - \epsilon^2 \frac{\pi^2 c}{48} \sum_{l,m} \left| f_{l,m} \right|^2 \frac{(l^2 - 1)(l+2)}{l} \left( \frac{ \Gamma\left( \frac{l+1}{2} \right) } { \Gamma\left( \frac{l}{2} \right) } \right)^2 + O(\epsilon^3),
\ee
with~$c$ the central charge of the theory, as given above.  To compute the vacuum energy for general~$\Sigma$ we must instead resort to numerical computations.  To make this more tractable we restrict to the case where~$\Sigma$ is axisymmetric; hence we take the metric to be
\be
\label{eq:deformedspheregeneral}
d\Sigma^2 = b(\theta)d\theta^2 + s(\theta)\sin^2{\theta} \, d\phi^2
\ee
(with the usual identification~$\phi \sim \phi + 2\pi$) with smoothness at the poles~$\theta = 0$,~$\pi$ requiring that the functions~$b(\theta)$ and~$s(\theta)$ be smooth in~$\theta$ there, as well as~$b(0) = s(0)$ and~$b(\pi) = s(\pi)$.

An appropriate transformation of~$\theta$ allows us to write~\eqref{eq:deformedspheregeneral} in a form conformal to the round sphere; hence the space of axisymmetric geometries in which we're interested is parametrized by a single function of~$\theta$.  This space is impossible to comprehensively explore numerically; we will therefore be satisfied with considering various one-parameter families of geometries which smoothly deform from the round sphere at~$\epsilon = 0$ to a singular geometry at some~$\eps \neq 0$.  We will focus on two classes of such geometries: \\

\begin{figure*}[t]
\centering
\captionsetup[subfloat]{justification=centering,labelformat=empty}
\subfloat[][$l = 2$ \\ $-0.51 < x < 0$]{
\includegraphics[width=0.17\textwidth]{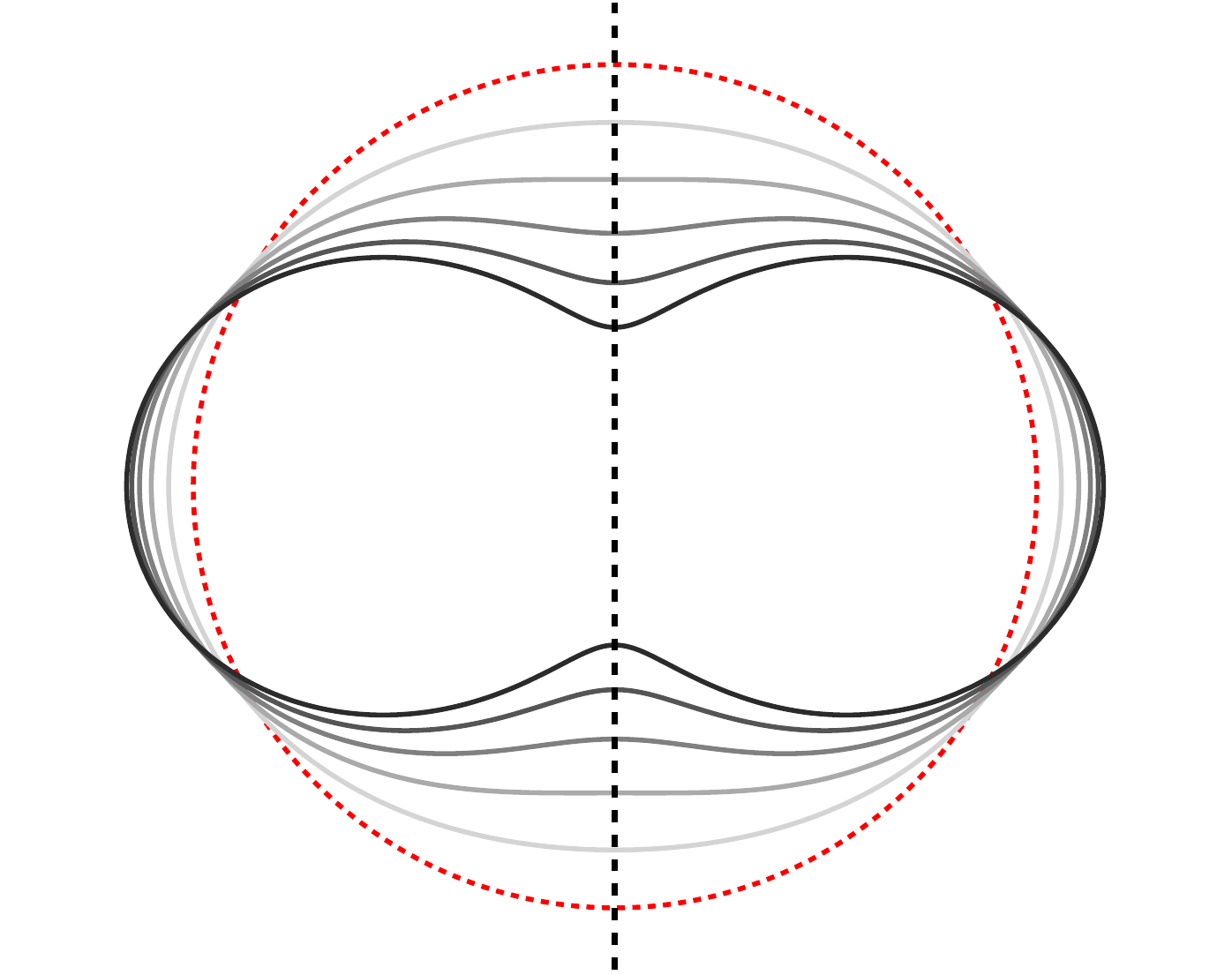}
}
\subfloat[][$l = 2$ \\ $0 < x < 0.6$]{
\includegraphics[width=0.1\textwidth]{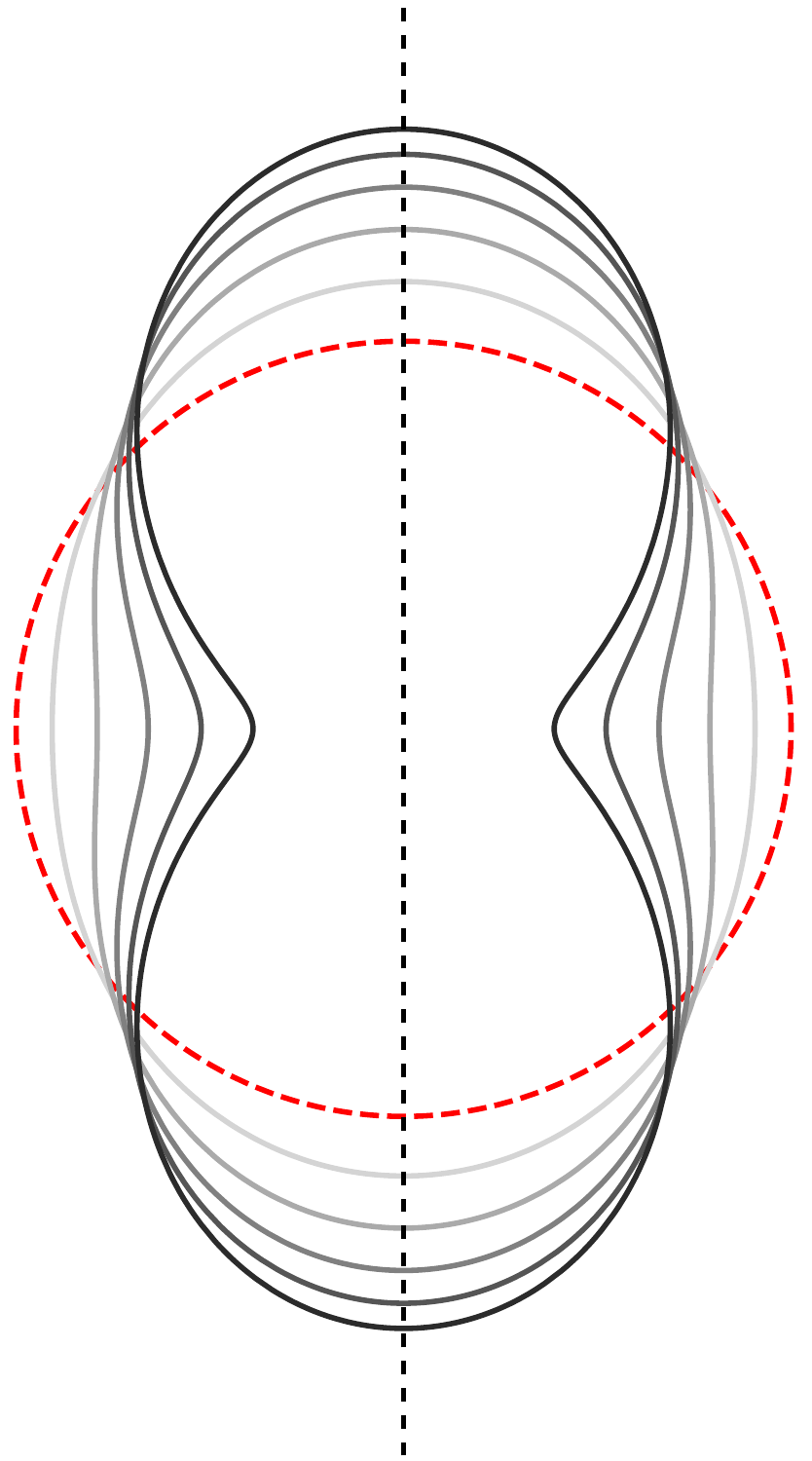}
}
\hspace{0.25cm}
\subfloat[][$l = 4$ \\ $-0.34 < x < 0$]{
\includegraphics[width=0.16\textwidth]{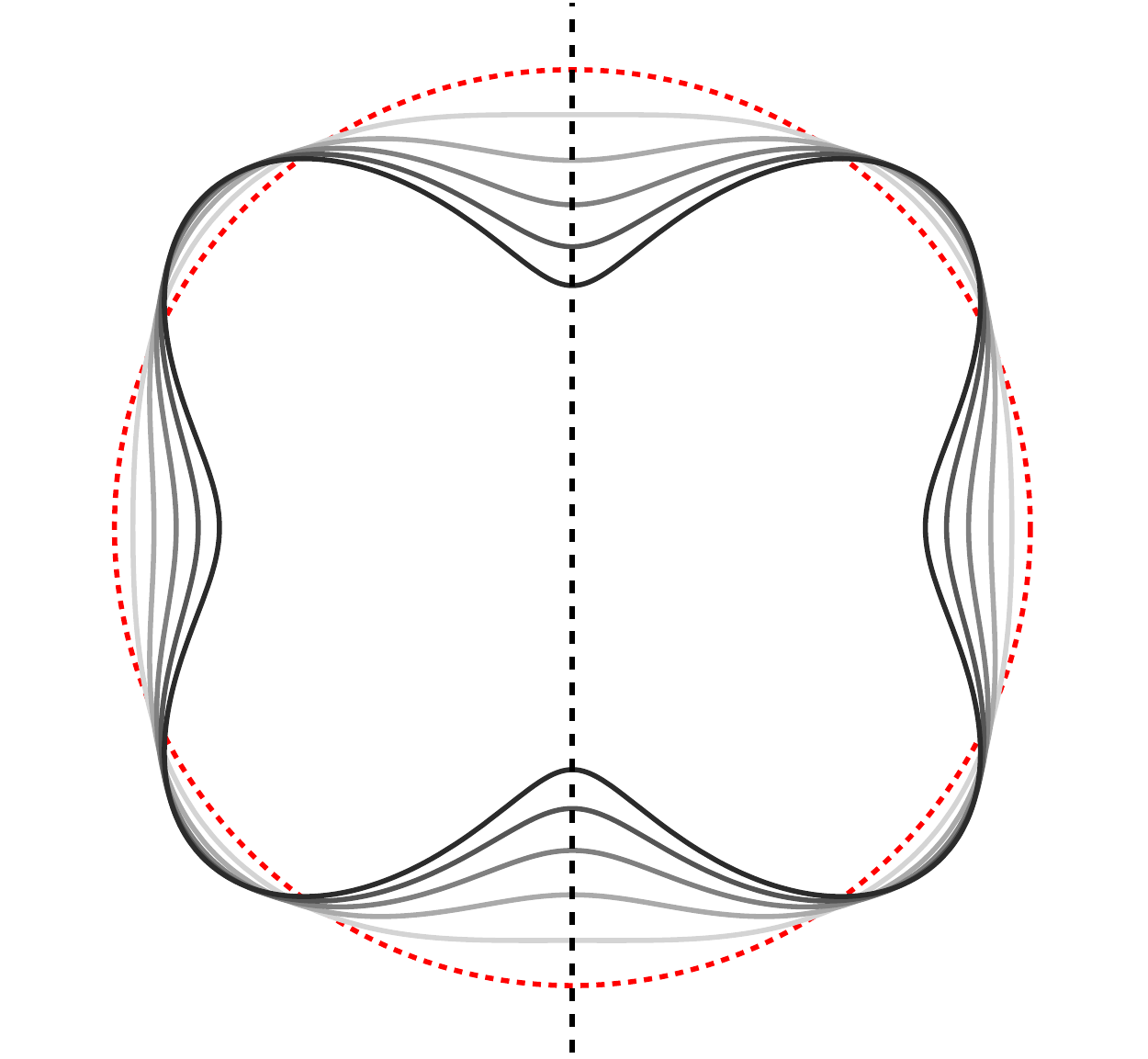}
}%
\subfloat[][$l = 4$ \\ $0 < x < 0.37$]{
\includegraphics[width=0.13\textwidth]{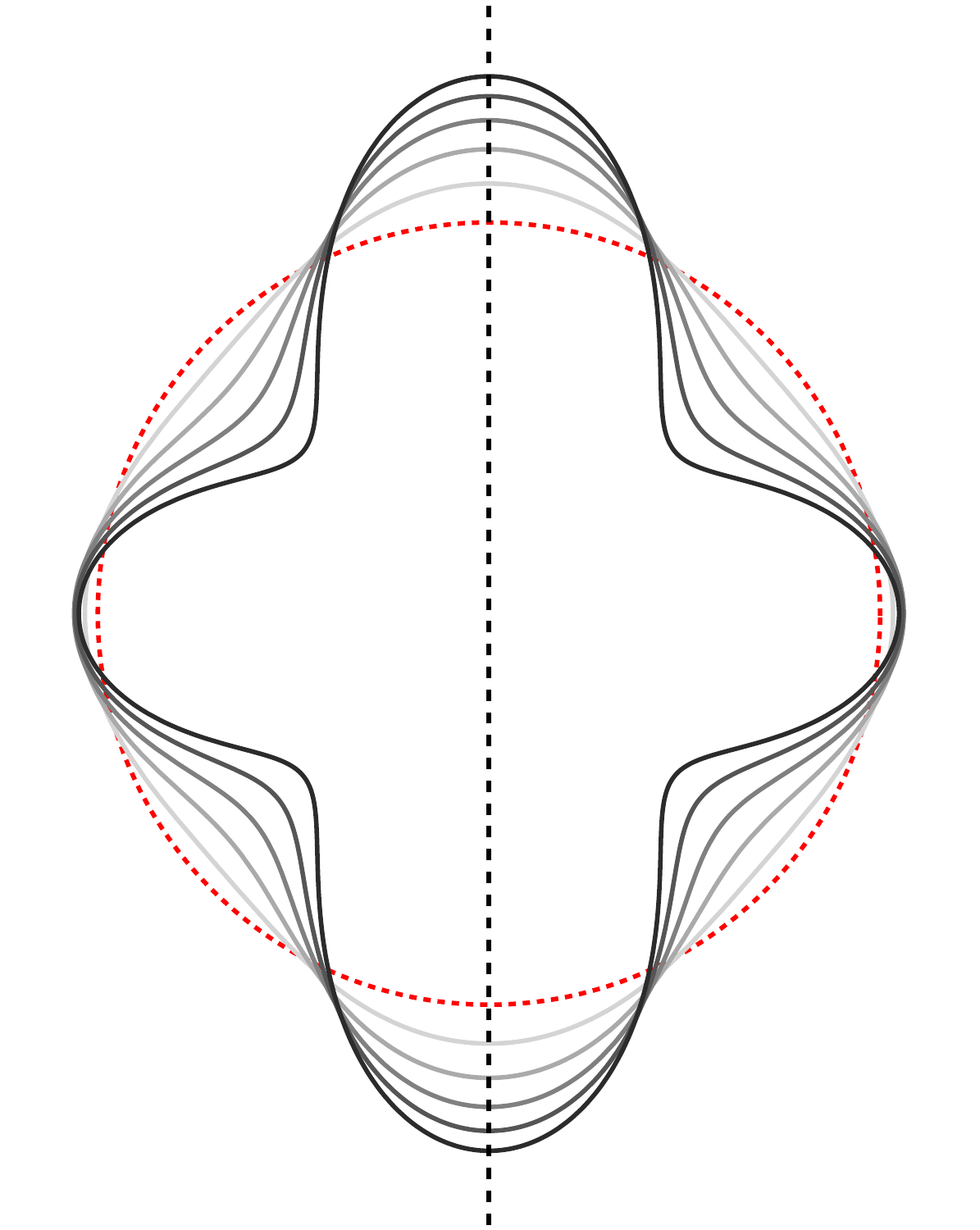}
}
\hspace{0.25cm}
\subfloat[][$l = 6$ \\ $-0.28 < x < 0$]{
\includegraphics[width=0.16\textwidth]{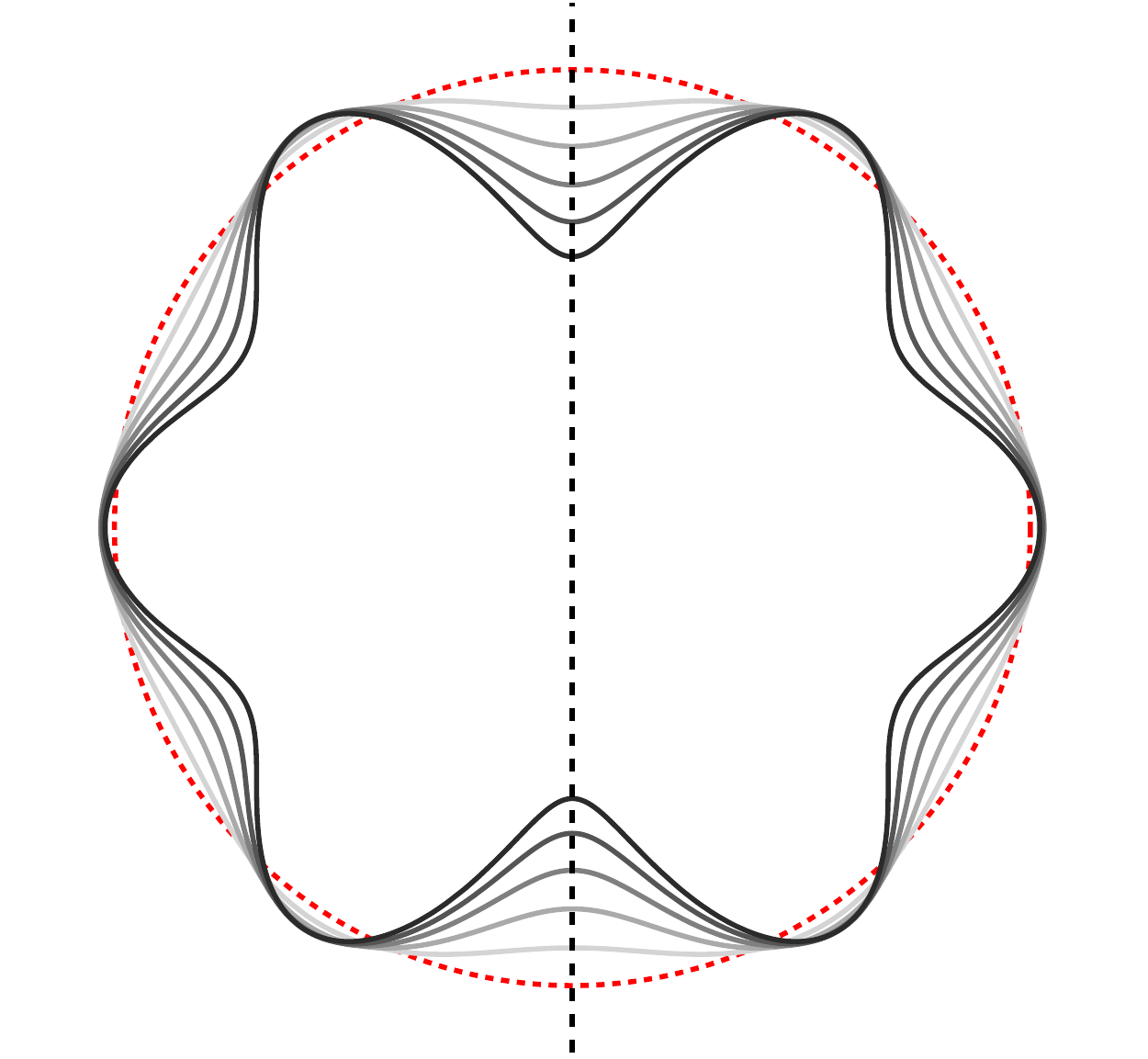}
}
\subfloat[][$l = 6$ \\ $0 < x < 0.3$]{
\includegraphics[width=0.13\textwidth]{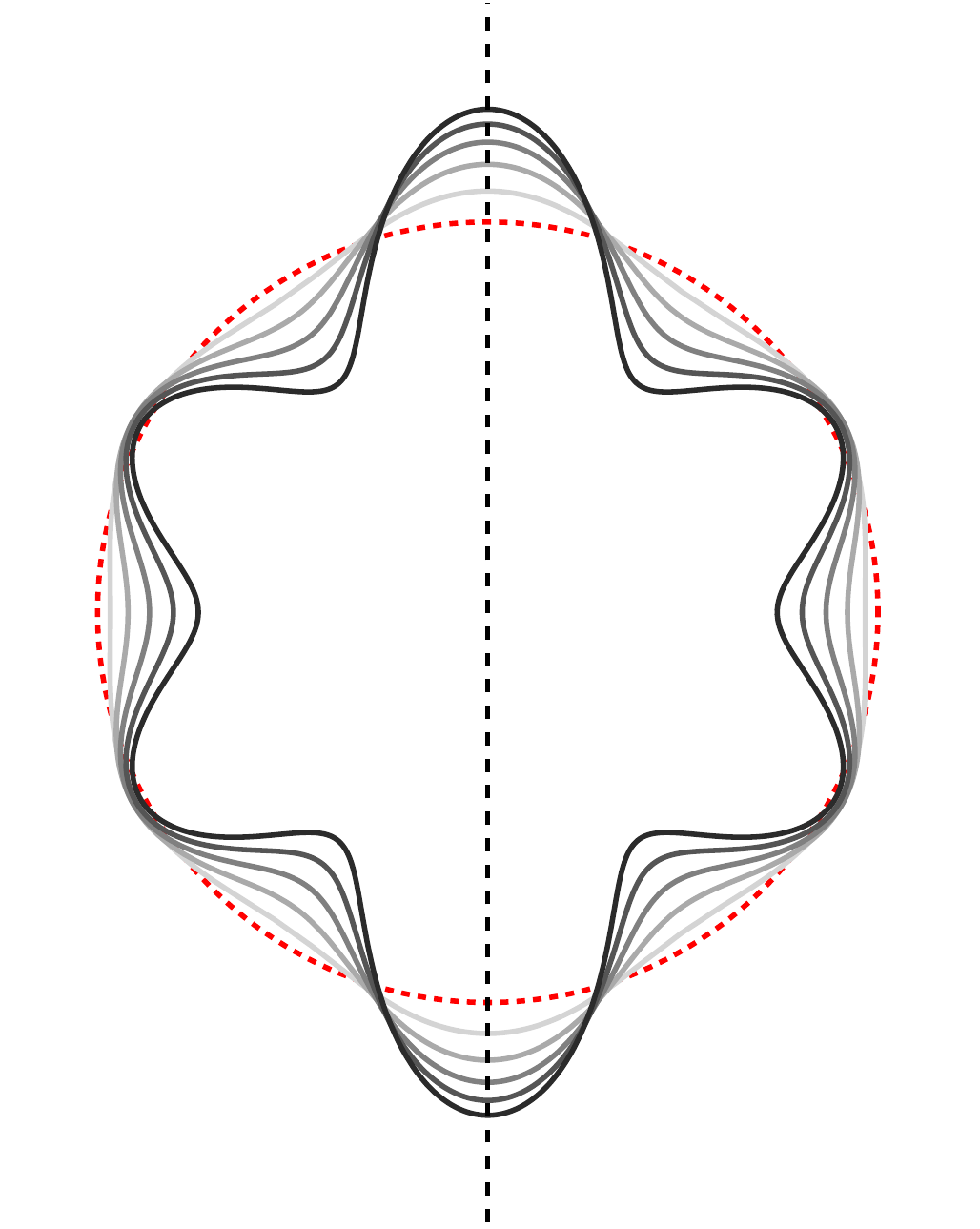}
}
\caption{Embeddings of the geometries~\eqref{eq:type1metric} in~$\mathbb{R}^3$ for~$l = 2$,~4, and~6; the full geometries are surfaces of revolution obtained by rotating about the dotted vertical line.  The red dotted circle represents the undeformed round sphere, while from light gray to black the other curves show even spacings in $x$ for the labeled range with~$x$ related to~$\eps$ as in~\eqref{eq:xdef}.  The black curves, corresponding to the extremal values of~$x$ for each~$l$, show the largest values of~$x$ for which we will present results in Section~\ref{sec:results}.}
\label{fig:type1embeddings} 
\end{figure*}

\noindent \textbf{Type 1 Geometries} are embedded in $\mathbb{R}^3$ as surfaces of revolution given by an embedding function~$r = R_\eps(\theta)$ in the usual $(r, \theta, \phi)$ spherical coordinates.  These embeddings lead to the metric
\be
\label{eq:type1metric}
d\Sigma^2 = (R_\eps(\theta)^2+R_\eps'(\theta)^2)d\theta^2 + R_\eps^2(\theta)\sin^2{\theta} \, d\phi^2 .
\ee
Specifically, we will consider the one-parameter families of embedding functions given by
\be
\label{eq:Rfn}
R_\eps(\theta) = a_{l,\eps} \left(1+\eps Y_{l,0}(\theta) \right)
\ee
for various~$l$, with~$a_{l,\eps}$ a constant chosen to keep the area fixed to~$4\pi$.  As we will briefly discuss in Section~\ref{sec:gravity}, restricting to parity-symmetric geometries allows us to reach greater accuracy in the holographic CFT computations.  Therefore we will make this restriction from now on, corresponding to considering only even~$l$.  For each~$l$, the range of~$\eps$ is bounded as~$\epsilon_{\textrm{min}} < \epsilon < \epsilon_{\textrm{max}}$ (arising from the requirement that~$R_\eps(\theta) > 0$ for all~$\theta$); as~$\eps \to \eps_\mathrm{min}$ the geometry remains connected but develops a cusp-like singularity, while as~$\eps \to \eps_\mathrm{max}$ the geometry pinches off into disconnected components.  In Figure~\ref{fig:type1embeddings} we plot these embeddings for the range of~$\eps$ for which we will present numerical results in Section~\ref{sec:results}. \\

\begin{figure}[t]
\centering
\captionsetup[subfloat]{justification=centering,labelformat=empty}
\subfloat[][$n = 1$ \\ $-0.96 \leq x \leq 1/7$]{
\includegraphics[width=0.17\textwidth]{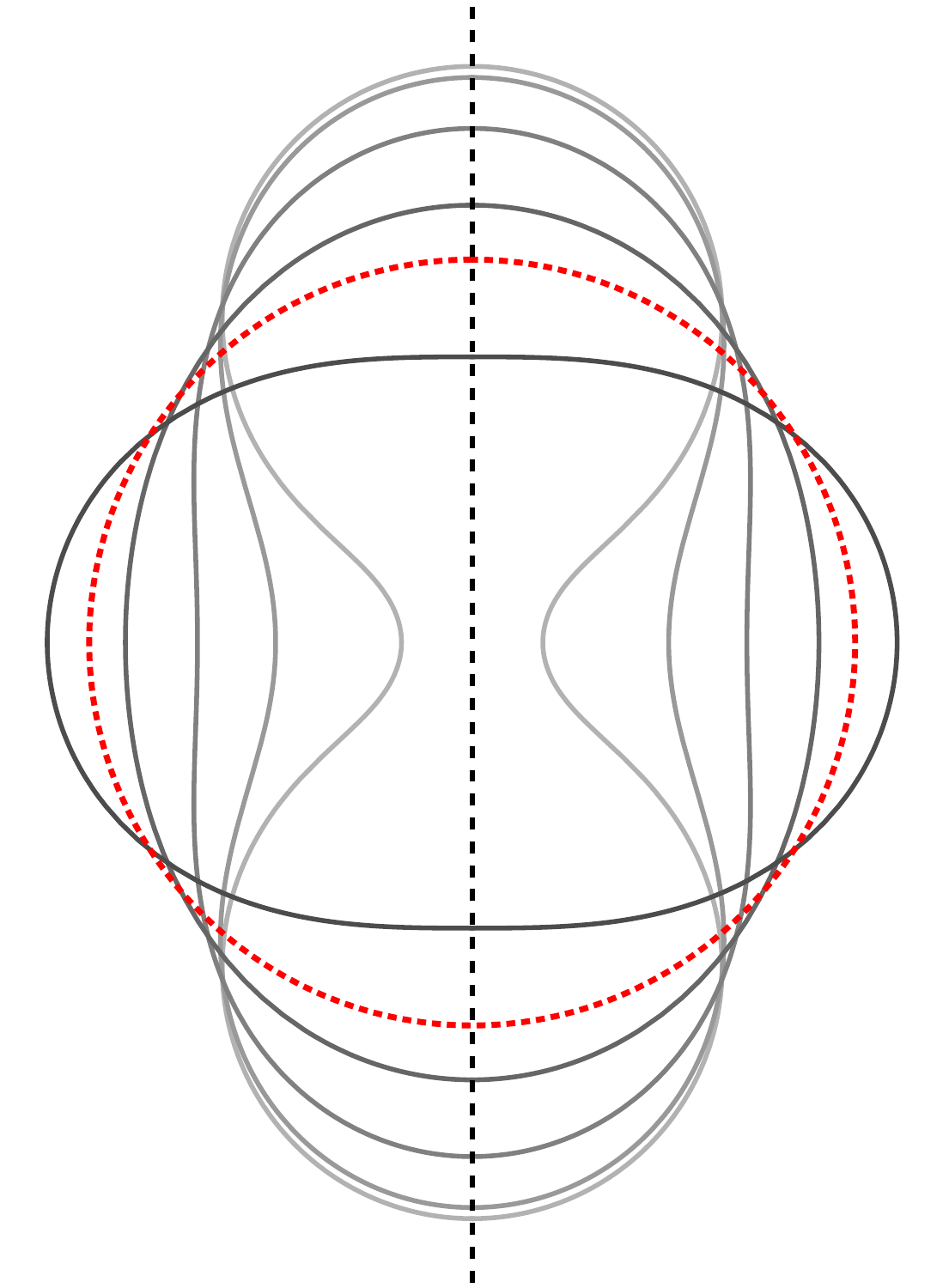}
}
\hspace{1cm}
\subfloat[][$n = 2$ \\ $-0.422 \leq x \leq 1/25$]{
\includegraphics[width=0.2\textwidth]{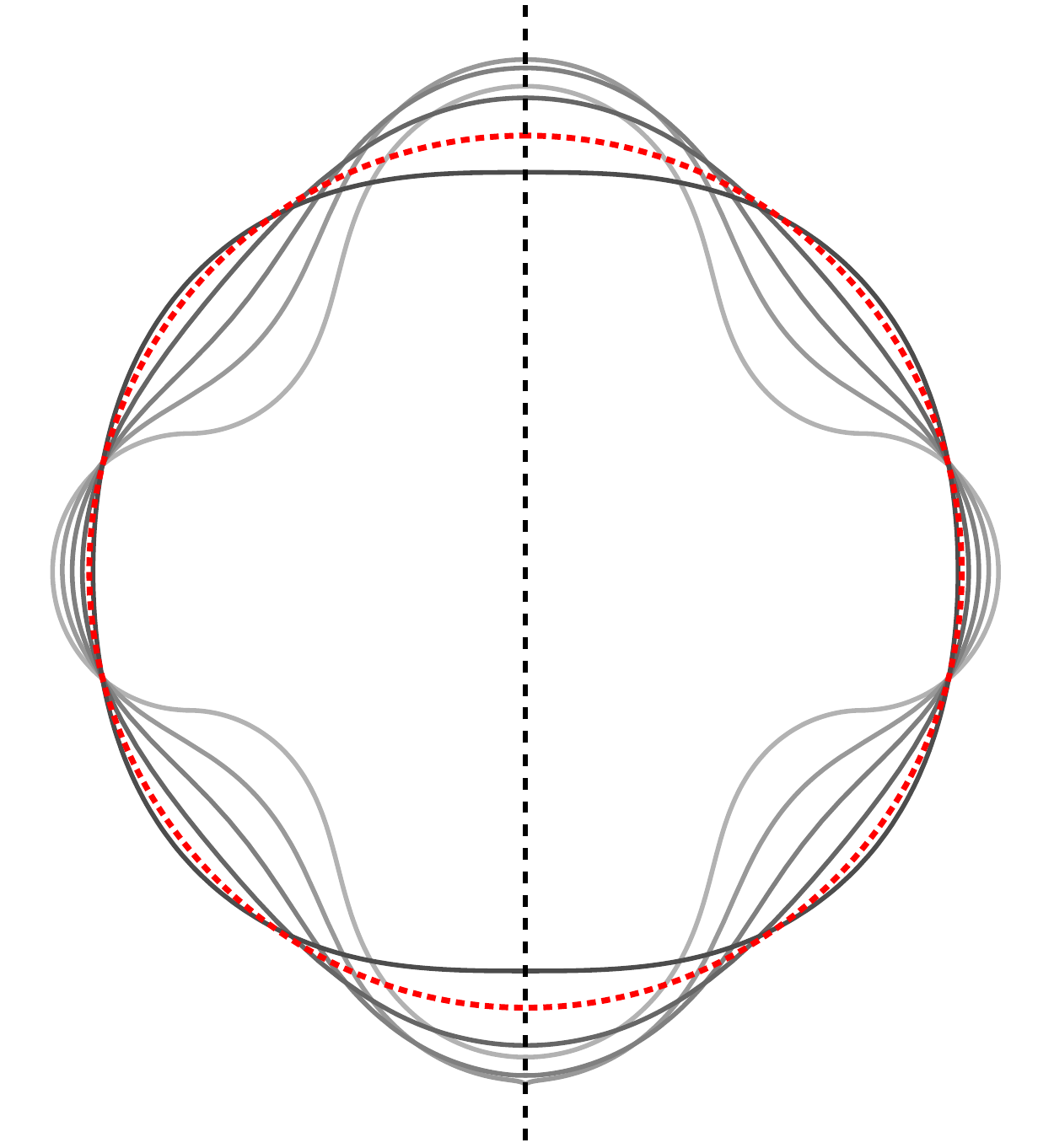}
}
\hspace{1cm}
\subfloat[][$n = 3$ \\ $-0.178 \leq x \leq 1/55$]{
\includegraphics[width=0.2\textwidth]{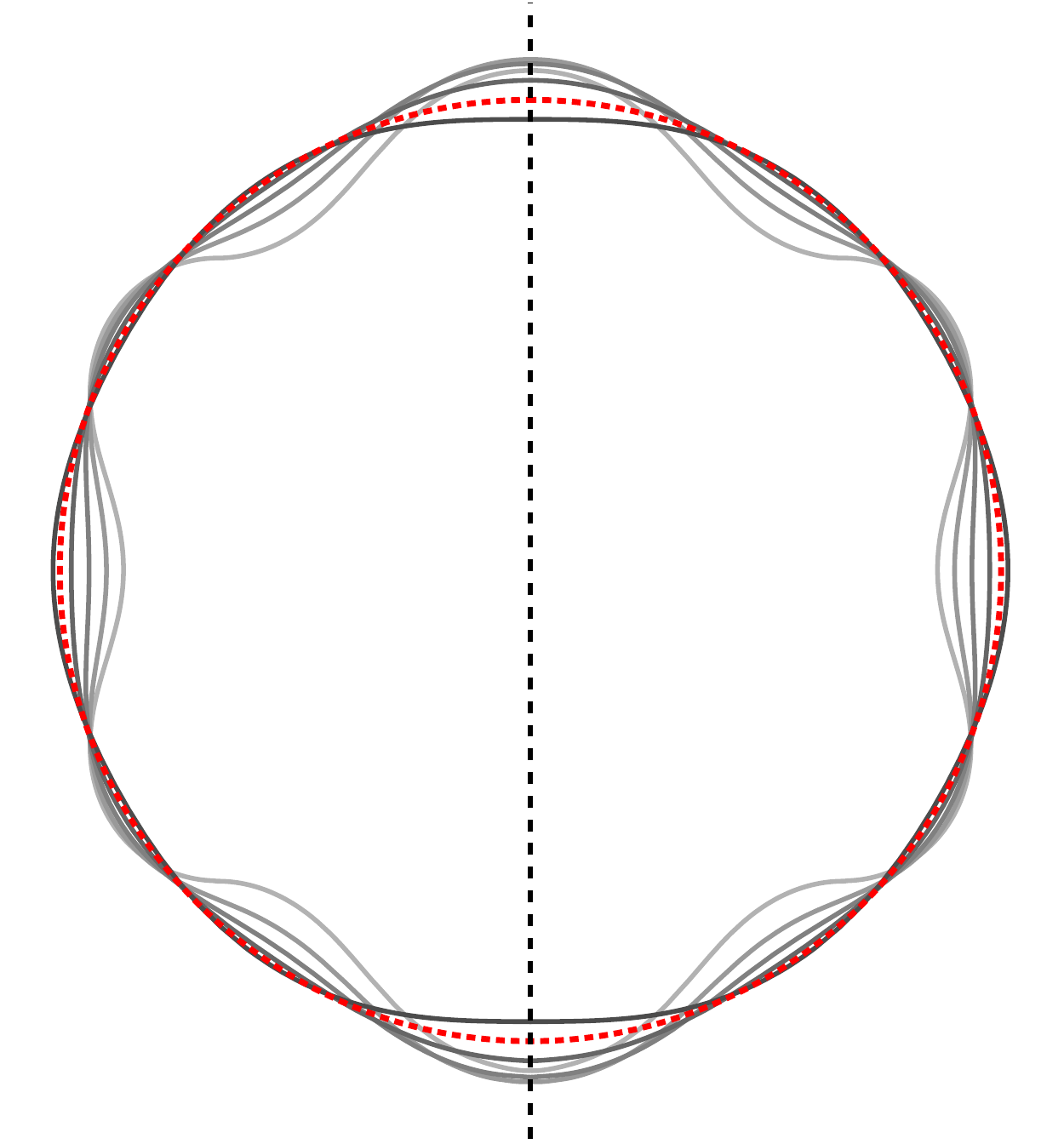}
}
\caption{Axisymmetric embeddings of the geometries~\eqref{eq:type2metric} in~$\mathbb{R}^3$ for~$n = 1$,~2, and~3 for the range of~$x$ for which embeddings exist; the full geometries are surfaces of revolution obtained by rotating about the dotted vertical line.  The red dotted circle represents the round sphere~$x = 0$, while the other curves show even spacings in~$x$; lines go from light to dark with increasing~$x$, with only the black lines corresponding to positive~$x$.  At positive~$x$, embeddings fail to exist for~$x > 1/(1+6n^2)$, correspondng to the Ricci curvature becoming negative at the poles (no axisymmetric embedding can have negative curvature at the poles).  For positive~$x$, corresponding to the lighter gray lines, the smallest values of~$x$ for which embeddings exist is determined numerically, except for the case~$n = 1$ for which embeddings exist all the way to~$x = -1$ when the sphere pinches off into two pieces (the lowest value~$x = -0.96$ shown here is the smallest value of~$x$ reached in the numerics discussed in Section~\ref{sec:results}).}
\label{fig:type2embeddings} 
\end{figure}

\noindent \textbf{Type 2 Geometries} are given by
\be
\label{eq:type2metric}
d\Sigma^2 =  \tilde{a}_{n,\epsilon} \left( d\theta^2 + H_{n,\eps}(\theta)\sin^2{\theta} \, d\phi^2 \right),
\ee
where now
\be
\label{eq:Hfn}
H_{n,\eps}(\theta) = 1 + \eps \sin^2\left(n \theta\right)
\ee
for various~$n$, and again $\tilde{a}_{n,\epsilon}$ is chosen to fix the area to be~$4\pi$.  These have even parity for~$n \in \mathbb{Z}$, and now~$\epsilon \in (-1, \infty)$.  Note that these geometries are not embeddable in~$\mathbb{R}^3$ for all~$n$ and~$\eps$; in Figure~\ref{fig:type2embeddings} we show those that are.   As~$\eps \to -1$ the geometry tends to pinch off into disconnected components, while as~$\eps \to \infty$ it remains connected but develops a cusp-like singularity. \\

The type~1 geometries~\eqref{eq:type1metric} are those previously explored in~\cite{Fischetti:2020knf} for the free scalar field and Dirac fermion.  We also consider the type~2 geometries~\eqref{eq:type2metric} partly for variety, but also partly to ensure that the phenomena we see are not specifically tied to considering geometries that are embeddable in~$\mathbb{R}^3$.  Moreover, our numerical methods perform slightly better on the type~2 geometries 
%
%
than on type~1 due to the simpler form of the metric functions.  We also note that for both types of geometry, it will be useful to introduce a modified deformation parameter
\be
\label{eq:xdef}
x = \frac{\eps}{A \eps + B},
\ee
with~$A$ and~$B$ chosen so that~$x \in (-1,1)$\footnote{Explicitly, for the type 1 geometries we take~$A = (\eps_\mathrm{max} + \eps_\mathrm{min})/(\eps_\mathrm{max} - \eps_\mathrm{min})$,~$B = -2\eps_\mathrm{min}\eps_\mathrm{max}/(\eps_\mathrm{max} - \eps_\mathrm{min})$, while for the type 2 geometries we take~$A = 1$,~$B = 2$.}.

Our primary purpose is to compute the vacuum energy of a holographic CFT for these classes of geometries, and to compare to the
%
results
obtained for the conformally coupled scalar and the massless free Dirac fermion.  These latter two CFTs have Euclidean actions given by
\begin{subequations}
\label{eqs:CFTactions}
\begin{align}
S_E[\phi] &= \frac{1}{2} \int d^3 x \sqrt{g} \, \phi \left(-\grad^2 + \frac{1}{8} R \right) \phi, \\
S_E[\bar{\psi},\psi] &= i \int d^3 x \sqrt{g} \, \bar{\psi} \slashed{D} \psi;
\end{align}
\end{subequations}
see e.g.~\cite{Fischetti:2018shp,Fischetti:2020knf} for more on our conventions.  The computation of the vacuum energy of these free theories is as described in~\cite{Fischetti:2020knf}.  In short, the vacuum energy can be expressed in terms of the heat kernel defined by the spatial part of the equations of motion of the actions~\eqref{eqs:CFTactions}.  Evaluating this heat kernel amounts to computing the spectrum of the spatial part of the equations of motion, which can be done numerically using standard pseudospectral methods.  We refer the reader to the original papers for more details on these computations, and to Appendix~\ref{subapp:free} where we give some details on the implementation and quantify its accuracy.  Our focus now turns to the evaluation of the vacuum energy of the holographic CFT.

%
\section{Holographic gravity solutions}
\label{sec:gravity}
%

Because we are considering CFTs dual to pure gravity, obtaining bulk solutions requires us to solve the vacuum Einstein equation with negative cosmological constant,
\be
\label{eq:Einstein}
R_{ab} = - \frac{3}{\ell^2} g_{ab},
\ee
for a static locally asymptotically AdS metric whose conformal boundary matches~\eqref{eq:deformedspheregeneral}.  Note that when the boundary sphere is deformed in an axisymmetric manner, we expect the static bulk to inherit this axisymmetry~\cite{AndChr02}, so we will restrict to axisymmetric bulk solutions.  For convenience we also now choose units in which the AdS length $\ell=1$. 

In numerically constructing the bulk solutions, we will follow the harmonic approach of~\cite{Headrick:2009pv,Figueras:2011va,Wiseman:2011by}. 
First, we write a general static and axisymmetric bulk metric as
\begin{multline}
\label{eq:AdSansatz}
ds^2 = \frac{1}{(1-r^2)^2}\bigg{[} -f(r)T(r, \theta)dt^2 + \frac{(1+r^2)^2}{f(r)}A(r,\theta)dr^2+2rF(r,\theta)\sin{\theta}\, dr \, d\theta \\
+ r^2\left( B(r,\theta)d\theta^2+S(r,\theta)\sin^2{\theta}\, d\phi^2\right) \bigg{]},
\end{multline}
where the functions~$\{ T, A, B, F, S \}$ depend only on the coordinate pair $(r, \theta)$,~$f(r) \equiv 1-r^2+r^4$, and since we are only looking at solutions that are even in parity under~$\theta \to \pi - \theta$, the coordinate domain can be taken to be~$r \in [0,1)$ and~$\theta \in [0,\pi/2]$, with~$r = 1$ corresponding to the AdS conformal boundary and~$r = 0$ to a bolt.  Note that with this ansatz, global AdS is obtained by setting~$T=A=B=S=1$ and $F=0$; this can be seen most easily by defining a new radial coordinate~$\rho = r/(1-r^2)$ in terms of which which the metric becomes
\be
ds^2_{\text{AdS}} = -(1+\rho^2)dt^2+\frac{d\rho^2}{1+\rho^2}+\rho^2d\Omega_2^2,
\ee
which is global AdS in standard global coordinates.

Instead of trying to solve~\eqref{eq:Einstein} directly for the various metric functions, we modify it into the so-called harmonic Einstein equation via the inclusion of an additional term:
\begin{equation}
\label{eq:EqnHarmonic}
R^H_{ab} \equiv R_{ab} - \nabla_{(a}\xi_{b)} + 3g_{ab} = 0,
\end{equation}
where the
%
DeTurck
 vector~$\xi^a$ is constructed from~$g_{ab}$ as well as a smooth reference metric $\bar{g}_{ab}$ and its Levi-Civita connection ${\Gamma^{a}}_{bc}[\bar{g}]$ via
\be
\xi^{a} = g^{bc}\left({\Gamma^{a}}_{bc}[g]-{\Gamma^{a}}_{bc}[\bar{g}]\right).
\ee
The reason for this modification is that~\eqref{eq:EqnHarmonic} evaluated on static geometries is an elliptic differential equation (in particular, its principal part is~$\mathrm{PP}\left[ R^H_{ab} \right] = - \frac{1}{2} g^{cd} \partial_c \partial_d g_{ab}$, which is simply a Riemannian Laplacian for each metric component when restricted to static geometries).  Of course, in order for a solution of~\eqref{eq:EqnHarmonic} to coincide with a solution to the original Einstein equation~\eqref{eq:Einstein}, we must have~$\grad_{(a} \xi_{b)} = 0$.  This is ensured by requiring that~$\xi^a = 0$, which can be thought of as fixing a gauge defined by the choice of reference metric~$\bar{g}_{ab}$.  Now, from~\eqref{eq:EqnHarmonic} it follows that~\cite{Figueras:2011va,Wiseman:2011by}
\be
\grad^2 \xi^2 + \xi^a \grad_a \xi^2 \ge 0
\ee
(with~$\xi^2 \equiv \xi^a \xi_a$), and hence the maximum principle for elliptic operators ensures that~$\xi^2$ can only have a maximum at the boundary of the domain, and that if it does attain such a maximum it must have positive outward normal gradient there.  Consequently, the vanishing of~$\xi^a$ everywhere can be enforced by choosing the reference metric~$\bar{g}_{ab}$ to obey the same boundary conditions as the desired solution~$g_{ab}$, ensuring that at each boundary of the domain either~$\xi^a$ or its normal gradient vanishes~\cite{Figueras:2011va}. 

These boundary conditions are as follows.  At the conformal boundary~$r = 1$, the metric must be conformal to the deformed sphere~\eqref{eq:deformedspheregeneral}, ensured by imposing the Dirichlet conditions
\be
T = A = 1, \quad F = 0, \quad B=b(\theta), \quad S = s(\theta)
\ee
there.

The coordinate boundary at~$r = 0$ corresponds to a bolt at which the (deformed) two-sphere shrinks to zero size. We may analyze the regularity conditions there by transforming to Cartesian coordinates $x = r \sin\theta \cos\phi$, $y = r \sin\theta \sin\phi$, and $z = r \cos\theta$ and requiring the metric components to be 
%
%
at least twice-differentiable (i.e.~$C^2$) functions of~$x$,~$y$ and~$z$.  
One finds the metric functions~$\{ T, A, B, F, S \}$ must be even in~$r$ and have an expansion in~$r$ which to leading order gives
\begin{subequations}
\label{eqs:smooth}
\begin{align}
T(r, \theta) &= T_0 + O(r^2) , \\ 
A(r, \theta) &= \frac{1}{2} \left( A_+ +  A_- \cos2\theta\right)  + O(r^2) , \\
B(r, \theta) &= \frac{1}{2} \left( A_+ -  A_- \cos2\theta\right)  + O(r^2) , \\
S(r, \theta) &= \frac{1}{2} \left( A_+ -  A_- \right)  + O(r^2) , \\
F(r, \theta) &=  -A_- \cos\theta  + O(r^2) 
\end{align}
\end{subequations}
for constants $T_0, A_\pm$. We include the points $r = 0$ in our domain and impose Neumann boundary conditions for the metric functions there.  Our initial guess for the metric and our choice of the reference metric both obey the smoothness conditions~\eqref{eqs:smooth}, and we implement a method of solution for the PDEs that preserves this smoothness in subsequent updates of the metric.  Thus any solution found should have the correct smooth behavior at the origin $r=0$.  In Appendix~\ref{subapp:holo} we use this smoothness as a diagnostic of the accuracy of our numerics.

The boundary at~$\theta = 0$ is the fixed point of the axisymmetry, and hence regularity there also requires the metric functions to be even in~$\theta$ and for~$B(\theta = 0) = S(\theta = 0)$; this is again ensured by taking Neumann boundary conditions at~$\theta = 0$ and having the initial guess and reference metric be smooth.  We also 
%
provide
checks of smoothness at this axis in the continuum limit in Appendix~\ref{subapp:holo}.

The parity symmetry~$\theta \to \pi-\theta$ requires the functions~$T$, $A$, $B$, and $S$ to be even about~$(\theta-\pi/2)$, while~$F$ should be odd there; hence at~$\theta = \pi/2$ we impose Neumann boundary conditions on~$T$, $A$, $B$, and $S$ and the Dirichlet condition~$F(\theta = \pi/2) = 0$.

With the boundary conditions specified, we take the reference metric to also be of the form~\eqref{eq:AdSansatz}, with metric functions given by
\bea
\overline{T} &= \overline{A} = 1, \quad \overline{F} = 0 \\
\overline{B} &= 1 + P(r) \left(b(\theta) - 1 \right), \\
\overline{S} &= 1 + P(r) \left( s(\theta) - 1 \right),
\eea
where~$P(r)$ is a function even in~$r$ that interpolates from zero at~$r=0$ to one at $r=1$; we take it to be
\be
P(r) = \frac{2 r^2}{1 + r^4}.
\ee
As required, this reference metric obeys all of the aforementioned boundary conditions (noting that~$b(\theta = 0) = s(\theta = 0)$ is required for regularity of the boundary metric~\eqref{eq:deformedspheregeneral})
%
and is at least twice-differentiable at~$r=0$ in the Cartesian coordinates~$(x,y,z)$ discussed above.
With this reference metric, it is straightforward to verify that~$\xi^2$ vanishes at the conformal boundary and has Neumann boundary conditions at the others (checked by expanding the harmonic equation~\eqref{eq:EqnHarmonic} about each boundary), and hence any solution of the harmonic equation~\eqref{eq:EqnHarmonic} must have~$\xi^a = 0$ everywhere by the arguments given above.

The system of elliptic PDEs obtained from~\eqref{eq:EqnHarmonic} is then solved numerically on the~$(r,\theta)$-domain by discretizing using pseudospectral differencing with Chebyshev-Gauss-Lobatto grid points in both directions.  This procedure involves a Newton-Raphson algorithm, for which we take the initial guess to be the reference metric~$\bar{g}_{ab}$.  We take the number~$N$ of grid points to be equal in both directions.  The results presented below were obtained with~$N = 60$, which yields very good accuracy provided the boundary metric is not close to being singular; indeed, typically our metric functions are accurate pointwise to one part in~$10^6$ or better.  A detailed discussion of the convergence and accuracy of our solutions is provided in Appendix~\ref{subapp:holo}.  We emphasize here, however, that our principal reason for restricting to parity-symmetric deformations of the round sphere is that we effectively double the grid size for free: the symmetry allows us to use~$N = 60$ points on the half-domain~$\theta \in [0, \pi/2]$ rather than in the full domain~$\theta \in [0,\pi]$.  It is certainly possible to consider non-parity symmetric deformations at the expense of extending the computational domain to the full domain of~$\theta$, but this substantially reduces the accuracy we are able to achieve for these gravitational solutions.  Importantly, the delicate comparison of the holographic CFT to the free fermion presented in Section~\ref{sec:results} below is much more difficult without the benefit afforded by parity symmetry.

Once a bulk solution has been obtained, our task is to compute the Casimir energy of the corresponding CFT state.  The most obvious approach would be to compute the boundary renormalized stress tensor, whose explicit expression is straightforward to obtain from an expansion of the bulk solution around~$r = 1$~\cite{BalKra99,deHSol00}.  From this one would take the $tt$-component, corresponding to the local CFT energy density, and integrate it over the spatial boundary geometry to obtain the total energy.  However, such an expression for the renormalized boundary stress tensor involves up to third 
%
derivatives
of the metric functions, and while in practice our solutions are highly accurate pointwise, this accuracy reduces dramatically if one takes several derivatives of these functions.  Consequently, extracting the boundary stress tensor in this manner leads to large systematic errors in the total energy at the resolutions used here.  It is also worth noting that the problem is compounded by large cancellations in the energy density, whose presence can be clearly inferred by considering small deformations of the boundary: in that case, the stress tensor varies at linear order in the deformation, whereas the total energy varies only quadratically due to these cancellations~\cite{FisHic16}.  For these reasons we introduce a novel method that is well-suited to highly accurate computation of the total energy. 

This method involves defining a so-called optical geometry~$\tilde{g}_{IJ}$ by decomposing the bulk metric as
\be
\label{eq:opticalmetric}
ds^2 = \frac{1}{Z(x)^2}\left(-dt^2 + \tilde{g}_{IJ}(x) dx^I \, dx^J\right);
\ee
here~$Z$ should be interpreted as a scalar field living on the optical geometry with~$Z = 0$ at the conformal boundary, so that~$\tilde{g}_{IJ}$ is regular in the conformal completion of the spacetime.  A particular conformal frame can be chosen by requiring that the restriction of~$\tilde{g}_{IJ}$ to the conformal boundary be the boundary metric~$h_{ij}$ in~\eqref{eq:metric}.  A useful property of the decomposition~\eqref{eq:opticalmetric} is that the bulk equations of motion imply that the Ricci curvature~$\widetilde{R}_{IJ}$ computed from~$\tilde{g}_{IJ}$ obeys~\cite{Boucher:1983cv,Galloway:2015ora,HicWis15}
\be
\label{eq:opticalRicci}
\widetilde{\nabla}^2 \widetilde{R} = -3 \left| \widetilde{R}_{IJ} - \frac{1}{3} \widetilde{R} \, \tilde{g}_{IJ} \right|^2 \leq 0,
\ee
where~$\widetilde{\grad}$ is the covariant derivative of the optical geometry~$\tilde{g}_{IJ}$.  Moreover, as shown in~\cite{HicWis15}, the boundary energy density can be expressed in terms of the normal gradient of~$\widetilde{R}$, so the Casimir energy can be expressed as
\be
E = \frac{c_h}{6} \int_{\Sigma} d^2y \sqrt{h}\, n^a \grad_a \widetilde{R},
\ee
where~$n^a$ is the unit outward pointing normal to the AdS boundary.  Now using the divergence theorem and~\eqref{eq:opticalRicci}, we may rewrite this expression for the Casimir energy as an integral over an entire time slice~$M$ of the bulk:
\be
\label{eq:Egrav}
E = - \frac{c_h}{2} \int_M d^3x \sqrt{\tilde{g}} \left| \widetilde{R}_{IJ} - \frac{1}{3} \widetilde{R} \, \tilde{g}_{IJ} \right|^2.
\ee
There are two important features of this rewriting.  First, we have reduced the number of derivatives we are required to take of the bulk metric from three to two.  Second, since the integrand is manifestly positive this expression does not suffer from cancellations.  As a result, computing this expression from the bulk metrics obtained numerically as described above gives a very accurate computation of the total energy.  It is worth also noting that this expression shows that the energy is manifestly non-positive, and in fact only vanishes for global AdS~\cite{HicWis15}.

%
\section{Results}
\label{sec:results}
%

We now compare the Casimir energy of the holographic CFT, computed from our numerical solutions using~\eqref{eq:Egrav}, to that of the conformally coupled scalar and of the free Dirac fermion.  Because the bulk geometries become more difficult to obtain as the boundary spheres approach becoming singular, we only present results for which we are confident in the holographic CFT calculation to about~$0.01$\% or better (the accuracy of the results we present is limited by the holographic computations; the Casimir energies of the free fields can be obtained comfortably to larger deformations).

\begin{figure}[t]
\centering
\includegraphics[width=0.62\textwidth]{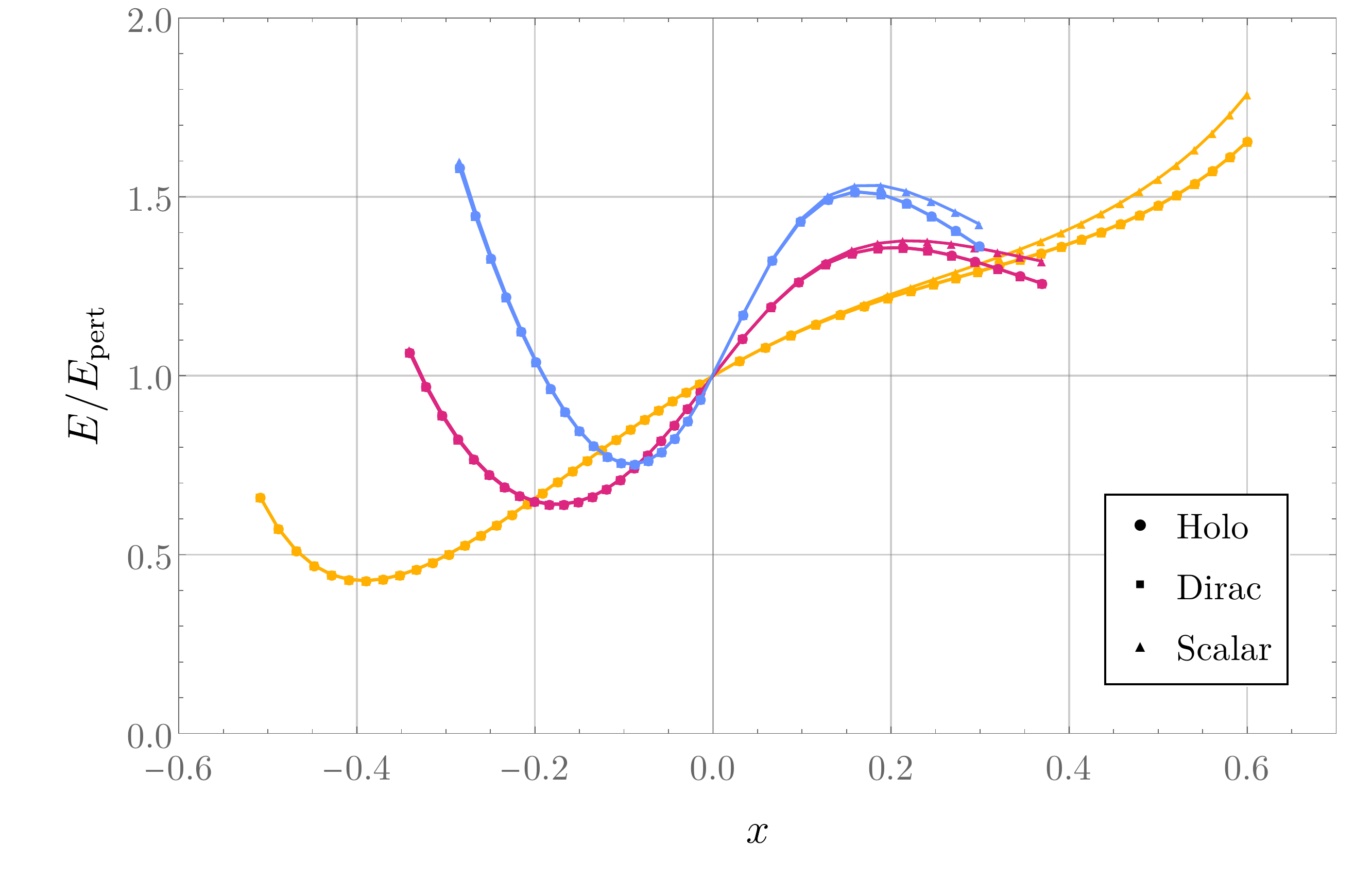}
\caption{The Casimir energy for the conformal scalar, Dirac fermion, and holographic CFT on the type~1 geometries~\eqref{eq:type1metric}; orange, magenta, and blue correspond to~$l = 2$,~4, and~6, respectively.  The energies are normalized by the perturbative behavior~\eqref{eq:Epert}, hence all the curves cross through~1 at~$x = 0$ by design.  Note that the data for the holographic CFT and the fermion cannot be distinguished by eye on this plot.  Also note that for visual clarity we show only the restricted domain~$x \in (-0.6,0.7)$ (all other plots in this paper show the full domain~$x \in (-1,1)$).}
\label{fig:type1energy}
\end{figure}

\begin{figure}[t]
\centering
\includegraphics[width=0.62\textwidth]{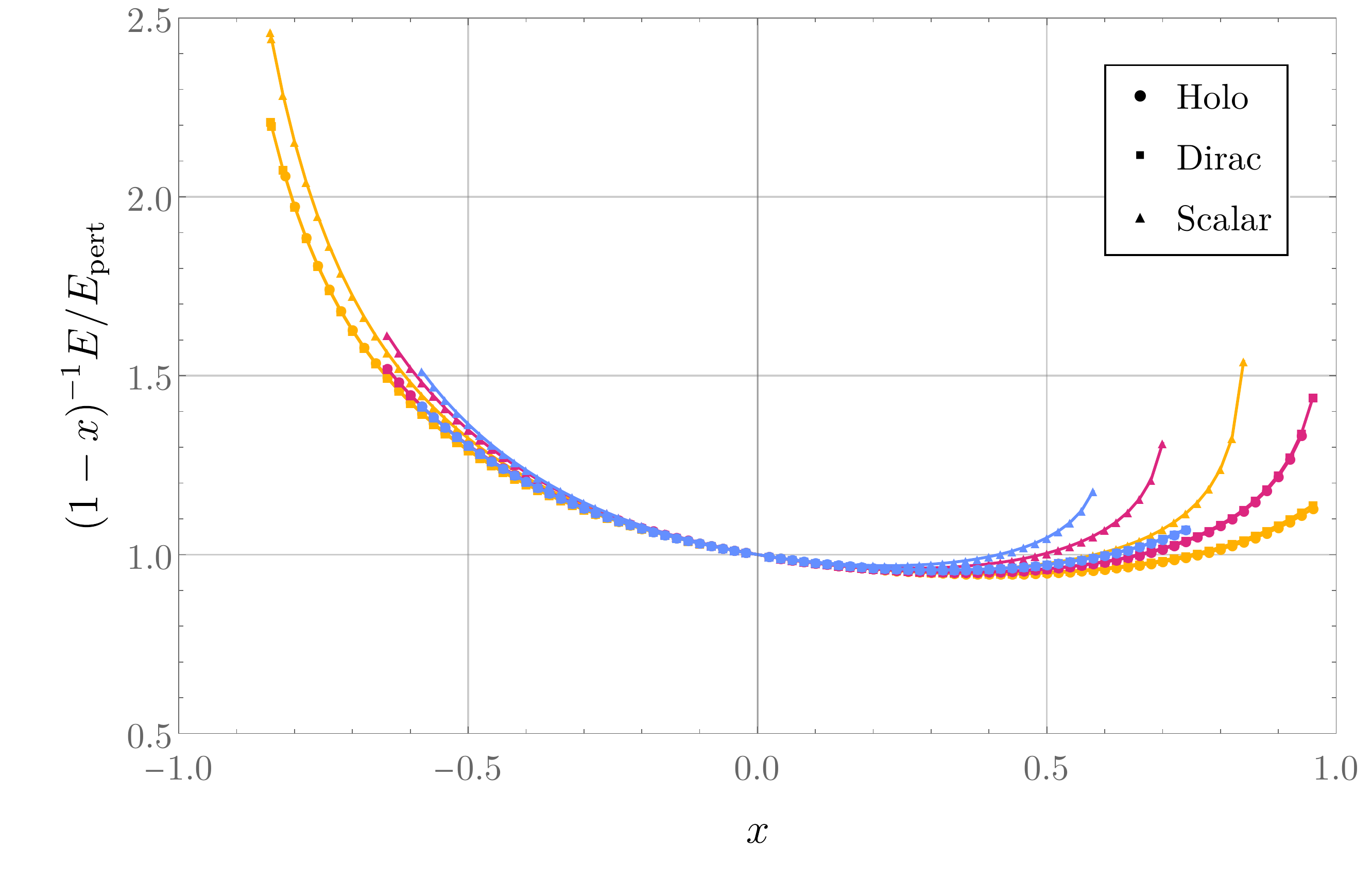}
\caption{The Casimir energy for the conformal scalar, Dirac fermion, and holographic CFT on the type~2 geometries~\eqref{eq:type2metric}; orange, magenta, and blue correspond to~$n = 1$,~2, and~3, respectively.  The energies are normalized by the perturbative behavior~\eqref{eq:Epert}, hence all the curves cross through~1 at~$x = 0$ by design; note also an additional normalization factor of~$(1-x)^{-1}$, introduced to clarify the behavior near~$x = 1$.  We emphasize that the data for the holographic CFT and the fermion cannot be distinguished by eye on this plot.
%
We note that the scalar theory has less extent in the positive~$x$ sense than the other theories as it ceases to exist when the conformal Laplacian~$-\grad^2 + R/8$ fails to be positive.
}
\label{fig:type2energy}
\end{figure}

\begin{figure}[t]
\centering
\includegraphics[width=0.6\textwidth]{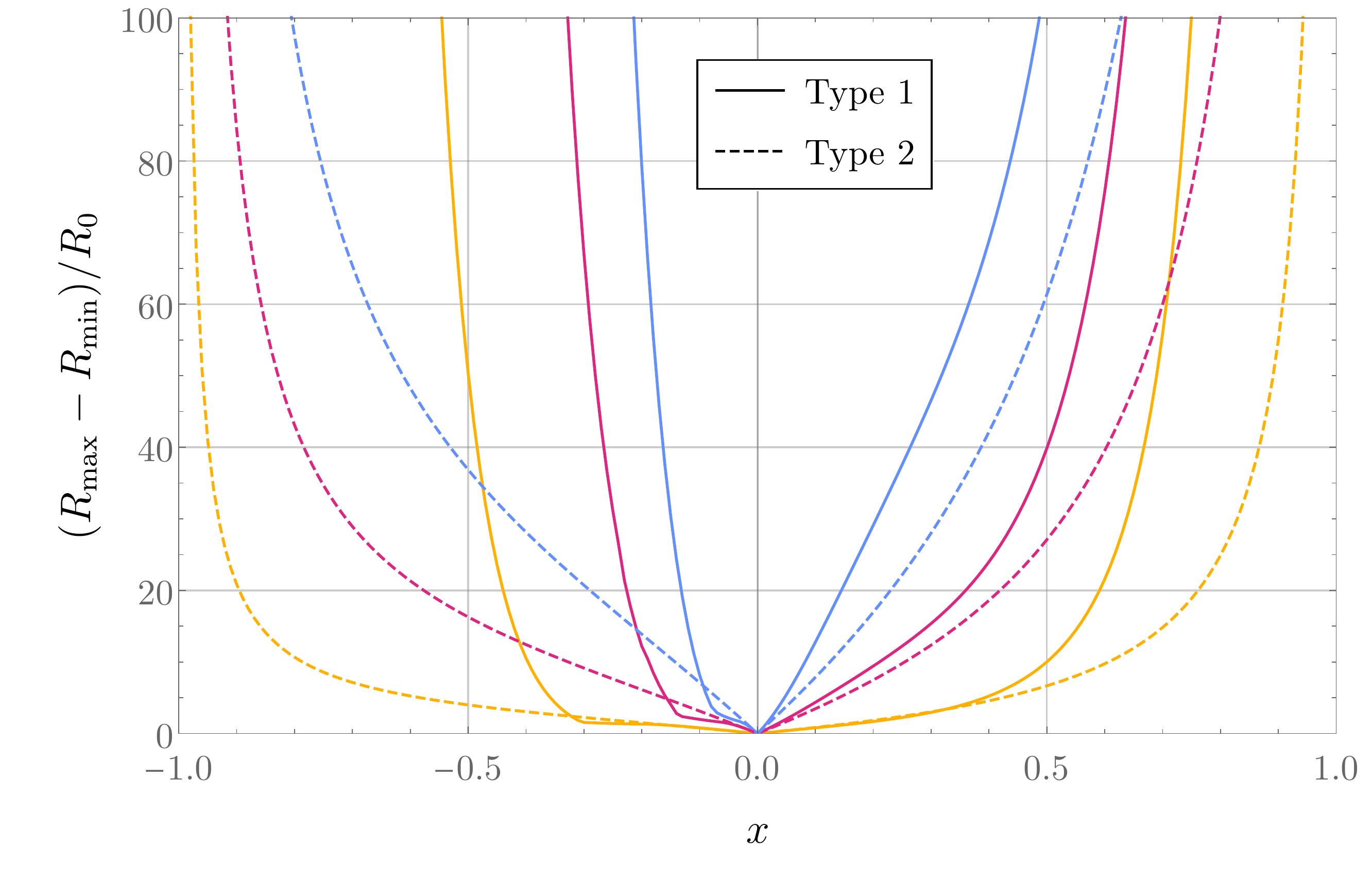}
\caption{The difference~$(R_\mathrm{max} - R_\mathrm{min})/R_0$ between the maximum and minimum values of the Ricci scalar on the geometries~\eqref{eq:type1metric} and~\eqref{eq:type2metric}, normalized by the value on the round sphere~$R_0 = 2$; solid orange, magenta, and blue correspond to the~$l = 2$,~4, and~6 Type 1 geometries, while dashed orange, magenta, and blue correspond to the~$n = 1$,~2, and~3 Type~2 geometries.}
\label{fig:Ricci}
\end{figure}

In Figure~\ref{fig:type1energy} we show the Casimir energy of the type~1 deformations for~$l = 2$,~4, and~6, normalized by the perturbative expectation~\eqref{eq:Epert}.  Outside the range of~$x$ for which we show data, either the gravity solutions have significant numerical error at the maximum resolutions we have used, or we are unable to find solutions at all.  However, we emphasize that bulk solutions likely still exist.  We see no obvious pathology in the bulk, and the difficulty in constructing the solutions is simply due to the boundary metric, and hence near-boundary bulk behaviour, becoming increasingly singular.
Likewise, in Figure~\ref{fig:type2energy} we show the Casimir energy of the type~2 deformations for~$n = 1$,~2, and~3, again normalized by the perturbative expectation~\eqref{eq:Epert}.  Note that in this case we are able to reach larger values of~$x$ than those shown in Figure~\ref{fig:type1energy}; in part this is simply because the type~2 deformations grow more slowly with~$x$ than the type~1 ones.  To illustrate this more explicitly, in Figure~\ref{fig:Ricci} we show the difference~$(R_\mathrm{max} - R_\mathrm{min})/R_0$ between the maximum and minimum Ricci scalar of the boundary geometries~\eqref{eq:type1metric} and~\eqref{eq:type2metric}, normalized by the value~$R_0 = 2$ for the round sphere.  An important feature to note is that over the range of~$x$ shown in Figures~\ref{fig:type1energy} and~\ref{fig:type2energy}, the Ricci scalar varies by about two orders of magnitude, so these deformations are far from the perturbative regime\footnote{That our most extreme deformations are well outside the perturbative regime can also be seen more intuitively but more qualitatively by noting that the embeddings shown in Figure~\ref{fig:type1embeddings} simply look like large deformations of the round sphere.}.

It is worth noting that for large deformations, the conformal scalar may develop a negative eigenvalue of~$-\grad^2 + R/8$, which renders the theory ill-defined. For the ranges of~$x$ shown here this occurs for the positive-$x$ type 2 geometries. Indeed, in Figure~\ref{fig:type2energy} the scalar energies extend to smaller (positive) values of~$x$ than those of the Dirac and holographic theories for precisely this reason.  An interesting point is that the results of~\cite{Ginoux} show that when the conformal Laplacian~$-\grad^2 + R/8$ has a negative eigenvalue, there does not exist a Weyl rescaling of the full~$(2+1)$-dimensional spacetime to another static spacetime with positive Ricci scalar.  But as far as we are aware, the only arguments for the existence of bulk gravity solutions in our setting require a static Weyl frame with positive boundary scalar curvature~\cite{Anderson:2002xb}.  Since here we are able to construct holographic bulk solutions beyond the regime in which the conformal scalar is well-defined, our results show that bulk solutions apparently continue to exist even when there is no conformal frame where the boundary metric has positive curvature.

\begin{figure}[t]
\centering
\subfloat[][Fermion]{
\includegraphics[width=0.5\textwidth]{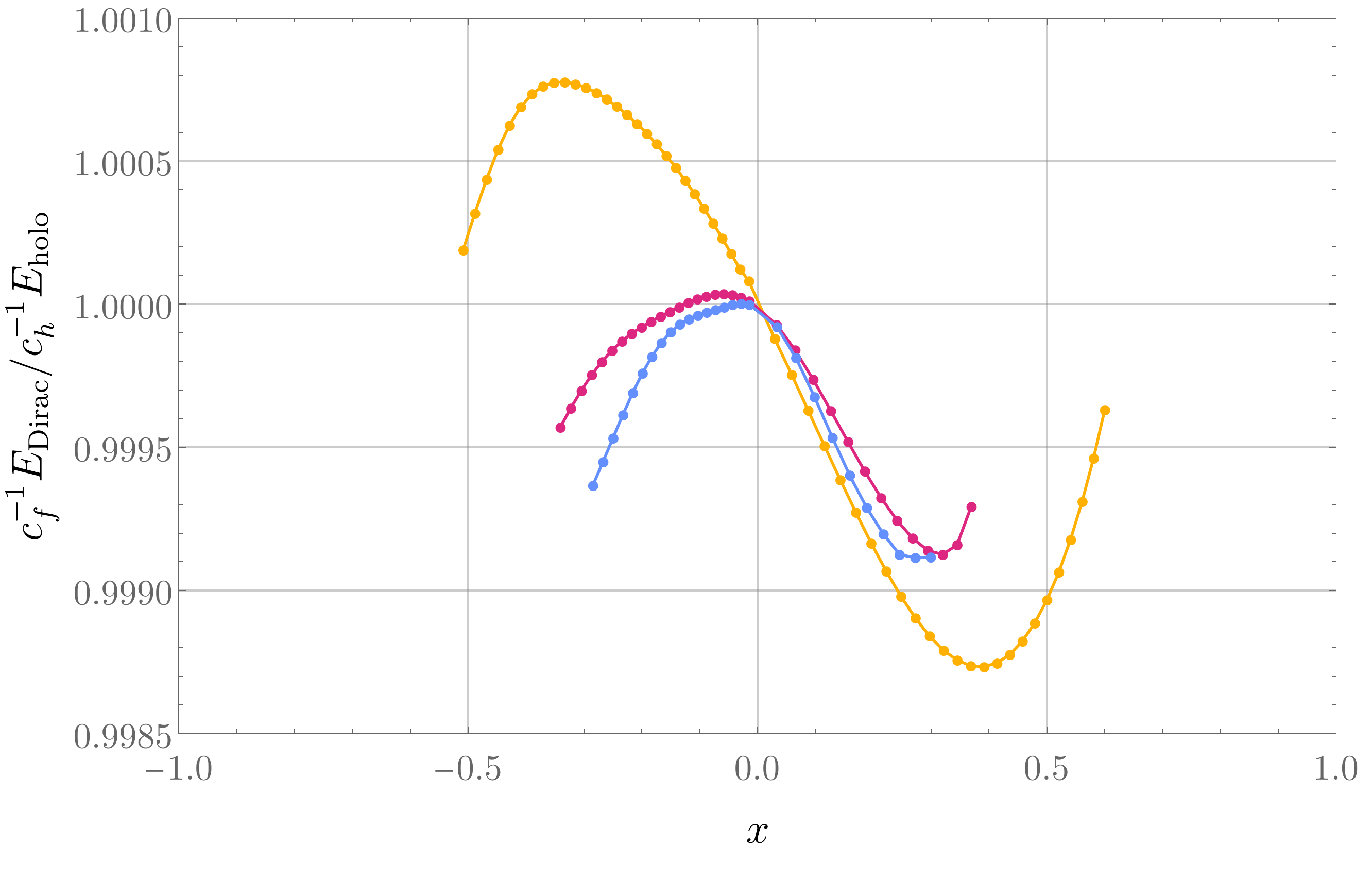}
\label{subfig:type1Diracholo}
}
\subfloat[][Scalar]{
\includegraphics[width=0.5\textwidth]{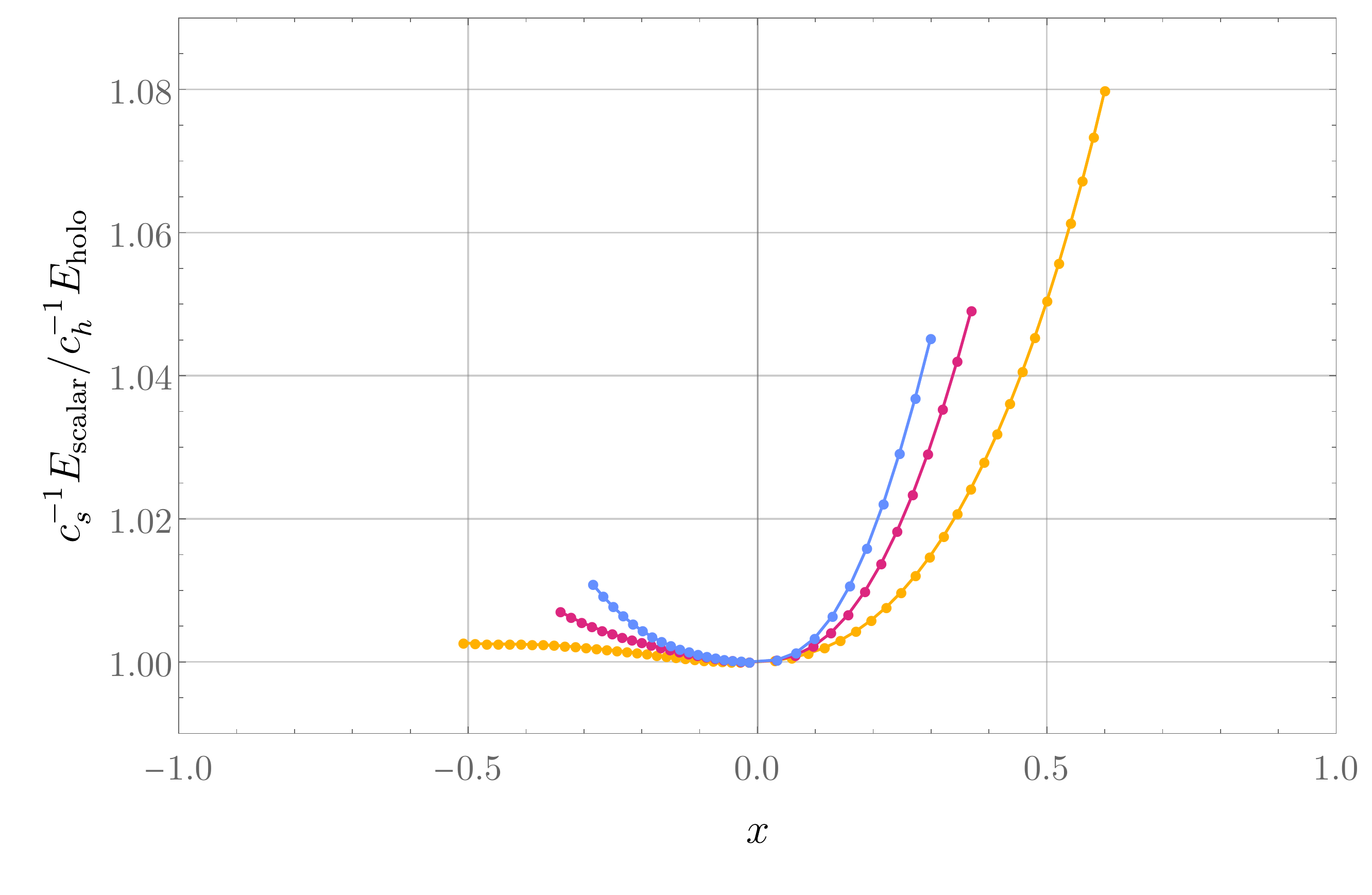}
\label{subfig:type1scalarholo}
}
\caption{The ratio between the Casimir energy of the holographic CFT and the free fields on the geometries~\eqref{eq:type1metric}; orange, magenta, and blue correspond to~$l = 2$,~4, and~6, respectively.  Note that the variation between the Dirac fermion and the holographic CFT is less than~$\sim 0.1$\% for all the deformations shown here, while the variation between the scalar and the holographic CFT reaches up to~$\sim 8$\%.}
\label{fig:type1ratios}
\end{figure}

\begin{figure}[t]
\centering
\subfloat[][Fermion]{
\includegraphics[width=0.5\textwidth]{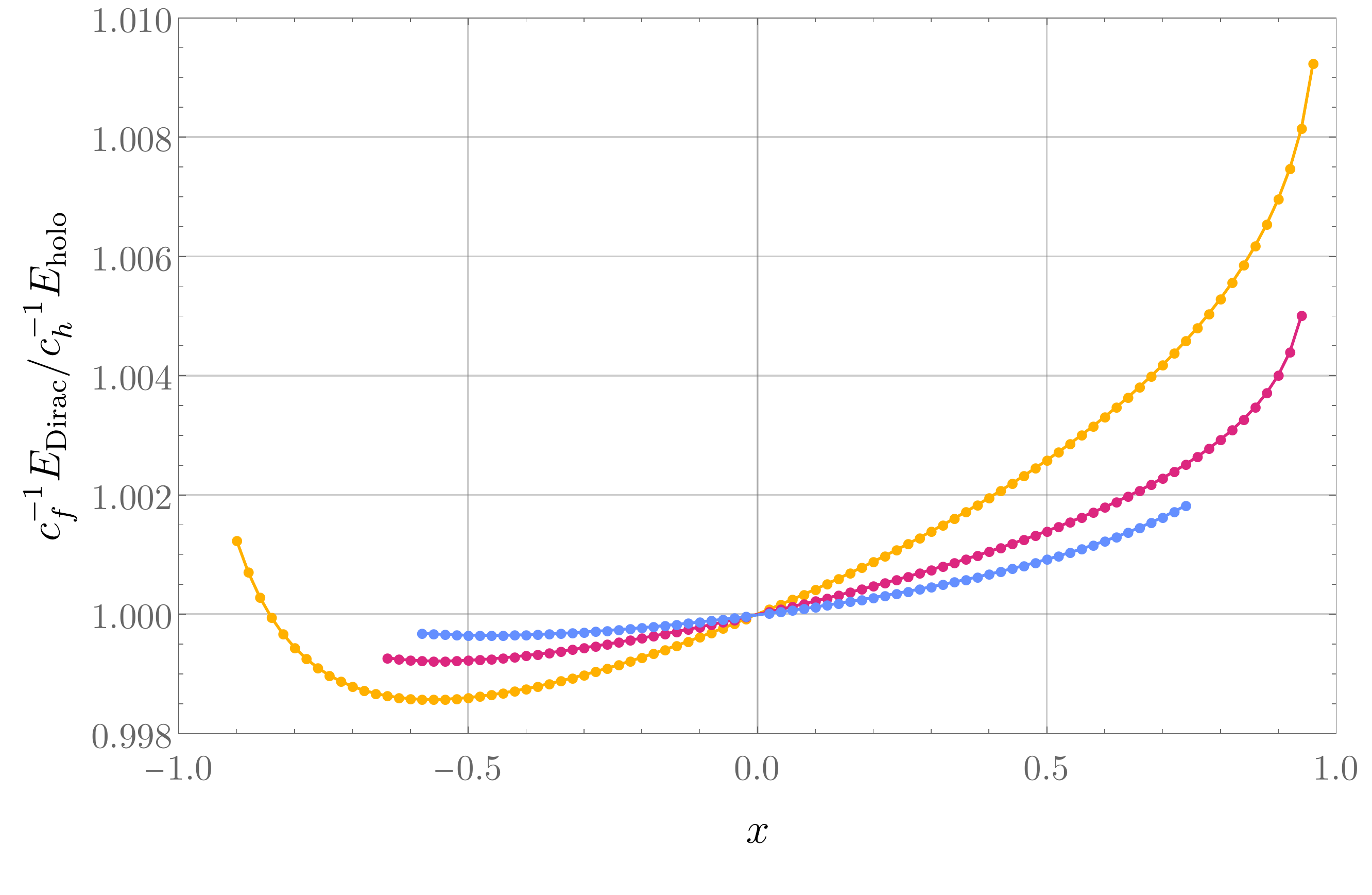}
\label{subfig:type2Diracholo}
}
\subfloat[][Scalar]{
\includegraphics[width=0.5\textwidth]{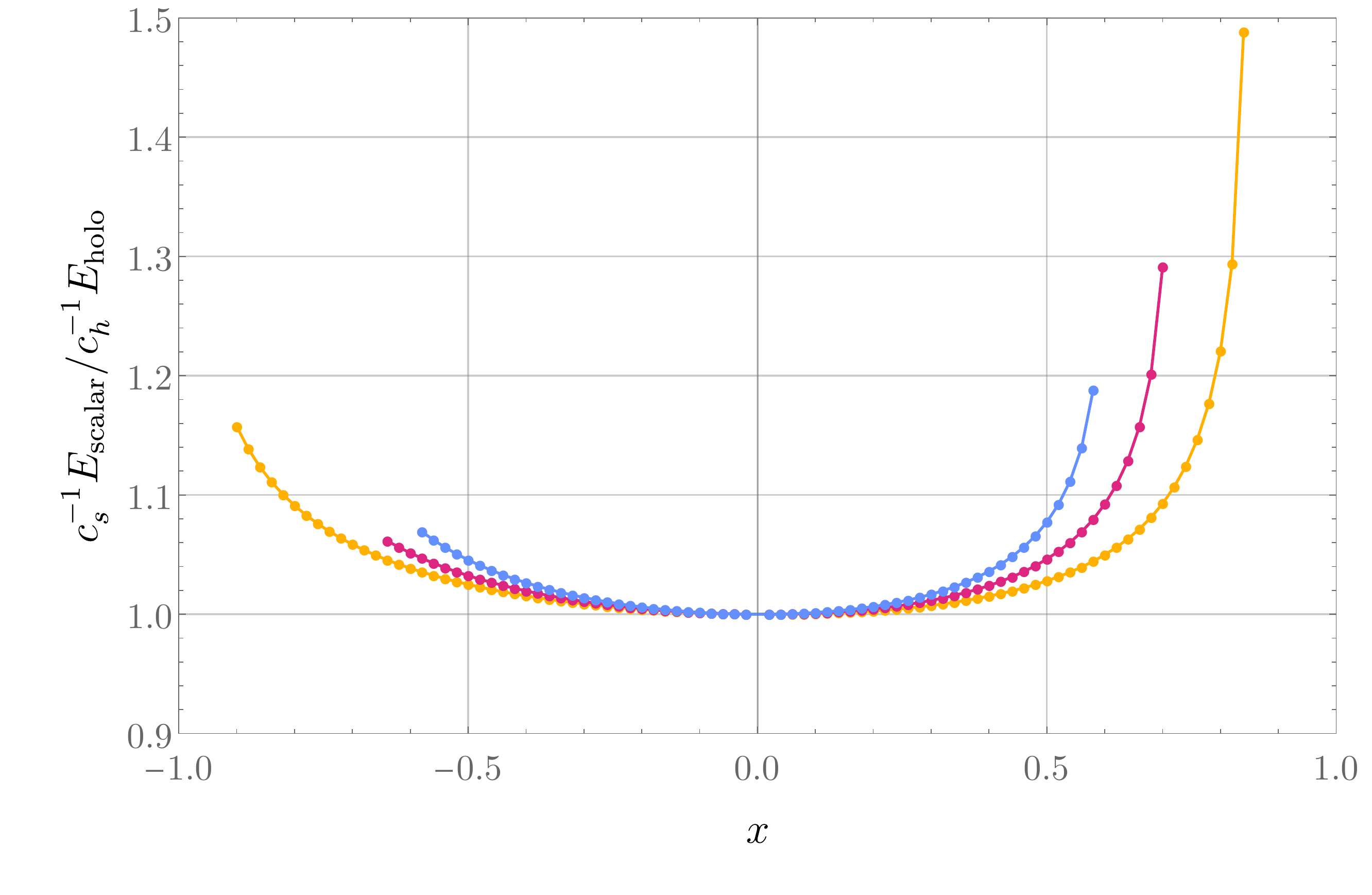}
\label{subfig:type2scalarholo}
}
\caption{The ratio between the Casimir energy of the holographic CFT and the free fields on the geometries~\eqref{eq:type2metric}; orange, magenta, and blue correspond to~$n = 1$,~2, and~3, respectively.  Note that the variation between the Dirac fermion and the holographic CFT is less than~$\sim 1$\% for all the deformations shown here, while the variation between the scalar and the holographic CFT reaches up to~$\sim 50$\% for the most extreme deformations.}
\label{fig:type2ratios}
\end{figure}

Returning to the vacuum energies, the first striking feature of Figures~\ref{fig:type1energy} and~\ref{fig:type2energy} is how similar the behavior of the Casimir energy is between the three theories.  This similarity had already been observed in~\cite{Fischetti:2020knf} between the fermion and the minimal scalar, even at nonzero temperature and mass (at high temperature, the similarity can be explained by a hydrodynamic expansion).  Here we see that the similarity extends to the conformal scalar and to the holographic CFT.  Perhaps more striking is the fact that the Dirac fermion and the holographic CFT almost perfectly coincide: in~\ref{subfig:type1Diracholo} and~\ref{subfig:type2Diracholo} we show the ratio~$c^{-1}_f E_\mathrm{Dirac}/c^{-1}_h E_\mathrm{holo}$ of the Casimir energies of the Dirac fermion and the holographic CFT normalized by their central charges, noting that they agree to~$\sim 0.1$\% for most of the deformations, and still to better than~1\% for the entire range we are able to accurately construct; this is reminiscent of the similarity found in~\cite{CheWal18} between the Dirac fermion and holographic CFTs on deformations of flat space, which persisted to nonzero temperatures.  Importantly, the difference shown in Figures~\ref{subfig:type1Diracholo} and~\ref{subfig:type2Diracholo} is substantially larger than any of our numerical uncertainties (which recall we restricted to be no larger than roughly~$0.01$\%), indicating that despite the close quantitative agreement, the fermion and the holographic CFT do not coincide \textit{exactly}.  We do not know of an explanation for this behavior; a putative explanation may well require some kind of fine-tuning. We also note that while we fix the area of the deformations to that of the undeformed sphere, this is irrelevant for the ratio of energies plotted in figures~\ref{subfig:type1Diracholo} and~\ref{subfig:type2Diracholo}, due to the scale invariance of the theories.

For comparison, in Figures~\ref{subfig:type1scalarholo} and~\ref{subfig:type2scalarholo} we also show the ratio~$c^{-1}_s E_\mathrm{scalar}/c^{-1}_h E_\mathrm{holo}$ of the Casmir energies of the conformally-coupled scalar and the holographic CFT normalized by their central charges; this ratio is much larger than that with the fermion.  Notably, for the Type~2 deformations shown in Figure~\ref{subfig:type2scalarholo} this ratio becomes quite large for the most extreme deformations at positive~$x$, indicating a substantial deviation between the behavior of the scalar field and the holographic CFT -- this deviation is of course related to the aforementioned fact that at large~$x$, the conformal scalar becomes ill-defined due to~$-\grad^2 + R/8$ acquiring a negative eigenvalue.

%
\section{Discussion}
\label{sec:disc}
%

We have numerically computed the vacuum energy of three different types of CFTs whose spatial geometries are given by the deformed spheres~\eqref{eq:type1metric} and~\eqref{eq:type2metric}, and we have seen that even for large deformations of the sphere the behavior of the free energy is remarkably qualitatively similar.  This similarity is perhaps even more surprising given that the scalar and Dirac fields are free theories, while the holographic theory is strongly-coupled.  Moreover, the agreement between the Dirac fermion and the holographic CFT is generally better than~0.1\%, and everywhere better than~1\%  for the deformations we can construct and over which we have good numerical control.  The strongest deviations are found for the most singular geometries where it becomes challenging to construct the gravity solutions, and also preserve good accuracy for the free theories.  Indeed, improving our numerical methods to allow an accurate determination of the Casimir energies closer to the singular limits of the geometries would be an interesting direction for future work.  Regardless of what happens for more extreme deformations, this close agreement between a free theory and a strongly-coupled one deserves a better understanding, especially in light of related results showing that the free massless fermion and the holographic CFT dual to Einstein gravity show close agreement in other ways, e.g.~in the entanglement entropy corner function~\cite{Bueno:2015rda,Bueno:2015xda} and in the ratio~$c_s/c$~\cite{Bueno:2015qya} mentioned in Section~\ref{sec:intro}.

One natural guess is that conformal invariance somehow enforces the behavior of the Casimir energy to be universal beyond the leading perturbative order~\eqref{eq:Epert}.  This is clearly not consistent with our results: the subleading behavior of the Casimir energy of perturbations of the round sphere is captured by the behavior of Figures~\ref{fig:type1ratios} and~\ref{fig:type2ratios} around~$x = 0$, and the nonzero slope there indicates that the Casimir energies of the three theories do not agree exactly beyond leading order.  Nevertheless, it is worth noting that in certain contexts there can indeed be close agreement between the free fermion, conformal scalar, and holographic CFTs dual to pure Einstein gravity beyond leading perturbative order.  To wit, the partition function of CFTs on the squashed Euclidean three-sphere~$S^3_\alpha$ has been studied in e.g.~\cite{Bobev:2017asb,Bueno:2018xqc,Bueno:2018yzo,Bueno:2020odt}:
\be
\label{eq:squashedsphere}
\ln Z[S_\alpha^3] - \ln Z[S^3_{\alpha = 0}] = c \left(a_2 \alpha^2 + a_3 \alpha^3\right) + \Ocal(\alpha^4),
\ee
where~$\alpha$ is a squashing parameter,~$c$ is the central charge,~$a_2$ is theory-independent due to conformal invariance, and~$a_3$ is constructed from the three-point function of the stress tensor on the round sphere.  In fact, for a large class of holographic CFTs~$a_3$ is proportional to the three-point function charge~$t_4$ (defined in~\cite{Hofman:2008ar}; see also~\cite{Osborn:1993cr}), while numerical results suggest that this proportionality also holds for the massless Dirac fermion and the conformal scalar~\cite{Bueno:2020odt}.  For holographic CFTs dual to Einstein gravity,~$a_3$ vanishes, while for the massless fermion and the conformal scalar it is nonzero but of order~$10^{-3}$.  Hence the subleading behavior of the squashed sphere partition function~\eqref{eq:squashedsphere} almost agrees for all three of these theories, reminiscent of the small but nonzero difference between the free energies of the holographic CFT and of the fermion shown in Figures~\ref{subfig:type1Diracholo} and~\ref{subfig:type2Diracholo}.

However, the results of~\cite{Bobev:2017asb} on the squashed three-sphere merely exhibited a surprising similarity between the partition functions of the scalar, fermion, and holographic CFT for \textit{small} squashings; for non-perturbatively large squashings, the fermion behaves very differently from the holographic CFT (in fact, the fermion partition function can even change sign relative to that of the unperturbed sphere, whereas that of the holographic CFT always has fixed sign).  Hence the close quantitative agreement between the Dirac fermion and the holographic CFT we have seen here must somehow rely on the particular structure of the deformations we have considered here.  One might therefore conclude that perhaps our results are merely an odd coincidence, potentially due to the symmetries we have imposed to make the computations numerically tractable; namely: ultrastaticity, axisymmetry, and parity.  The results of~\cite{Bobev:2017asb} confirm that breaking ultrastaticity, as the Euclidean three-sphere does, is enough to break the close matching between the fermion and the holographic CFT.  Why?  And what happens under the breaking of parity symmetry or axisymmetry?  It would be informative to check whether such deformations exhibit Casimir energies with similar behavior.  We leave such investigations as directions of future work.

%
\section*{Acknowledgements}
%

We would like to thank Simon Caron-Huot and Keivan Namjou for useful discussions and comments, and Pablo Bueno for comments on a previous version of this manuscript. Numerical calculations were performed on the Imperial HPC Research Computing
%
Facility.
The work was supported by the STFC grants ST/P000762/1 and ST/T000791/1, and LW by an STFC studentship. KC is sponsored by the DPST scholarship from the Royal Thai Government. SF is supported in part by the Simons Foundation Grant No.~385602 and the Natural Sciences and Engineering Research Council of Canada (NSERC), funding reference number SAPIN/00032-2015.

%
\appendix
\section{Numerical Details}
\label{app:convergence}
%

In this Appendix we include more details on the numerics, discussing specifically the heat kernel approach to computing the vacuum energies of the free theories, and commenting on the accuracy of our numerical computations.

\subsection{Free Theories}
\label{subapp:free}

For the free scalar and fermion, the Casimir energy is computed from heat kernels as described in~\cite{Fischetti:2020knf}.  In short, because for a CFT the renormalized Casimir energy of the round sphere vanishes, the renormalized Casimir energy of the deformed sphere can be defined as a difference between the energies of the deformed sphere and the round sphere.  This difference is given by
\be
\label{eq:heatkernel}
E[\Sigma] = \frac{\sigma}{\sqrt{4\pi}} \int_0^\infty \frac{dt}{t^{3/2}} \left[\Tr e^{-t L} - \Tr e^{-t \overline{L}}\right],
\ee
where~$\sigma = -1/2$ for the scalar and~$+1$ for the fermion,~$L$ is an elliptic differential operator on~$\Sigma$, and~$\overline{L}$ is the same differential operator on the round sphere.  To obtain the Casimir energy, we therefore need to compute the spectrum of~$L$ numerically and then perform the above integral (the spectrum of~$\overline{L}$ is known).  For the metrics~\eqref{eq:type1metric} and~\eqref{eq:type2metric}, the lowest-lying eigenvalues in the spectrum of~$L$ are obtained by exploiting the axisymmetry and using standard pseudospectral methods on a grid of~800 points in~$\theta$, which allows us to approximate the above traces by summing over the first~$\gtrsim 10^5$ eigenvalues.  Performing the integral over~$t$ is more subtle, as the traces in~\eqref{eq:heatkernel} do not commute with the integral.  Moreover, the small-$t$ behavior of the integrand is sensitive to the contributions of many eigenvalues of~$L$; hence for any approximation in which we truncate to the lowest-lying eigenvalues of~$L$, we are not able to accurately evaluate the integrand of~\eqref{eq:heatkernel} all the way to~$t = 0$.  Instead, we will make use of the heat kernel expansion to approximate the behavior of the traces at small~$t$~\cite{Vas03,Fischetti:2020knf}:
\be
\label{eq:smalltK}
\Tr e^{-t L} - \Tr e^{-t \overline{L}} = a t \int d^2 y \, \sqrt{h} \left(R - \overline{R}\right)^2 + \Ocal(t^2),
\ee
where~$R$ and~$\overline{R}$ are the Ricci scalars of the deformed and round spheres, and~$a$ is a theory-dependent (but geometry-independent) constant.  We then evaluate~\eqref{eq:heatkernel} by introducing a cutoff~$t_*$: for~$t > t_*$ we directly integrate the integrand of~\eqref{eq:heatkernel} using the (many) numerically-calculated lowest-lying eigenvalues of~$L$, but for~$t < t_*$ we instead integrate the expected linear small-$t$ behavior~\eqref{eq:smalltK}:
\be
\label{eq:truncatedE}
E_{t_*} \equiv \frac{\sigma}{\sqrt{4\pi}} \left[2 a \sqrt{t_*} \int d^2 y \, \sqrt{h} \left(R - \overline{R}\right)^2 + \int_{t_*}^\infty \frac{dt}{t^{3/2}} \left(\Tr_N e^{-t L} - \Tr_N e^{-t \overline{L}} \right) \right],
\ee
where the subscripts on the traces indicate that they are approximated by using only the lowest eigenvalues of~$L$, as described above. The above procedure was implemented using MATLAB.

\begin{figure*}[t]
\centering
\subfloat[][Type 1, $l = 6$]{
\includegraphics[width=0.5\textwidth]{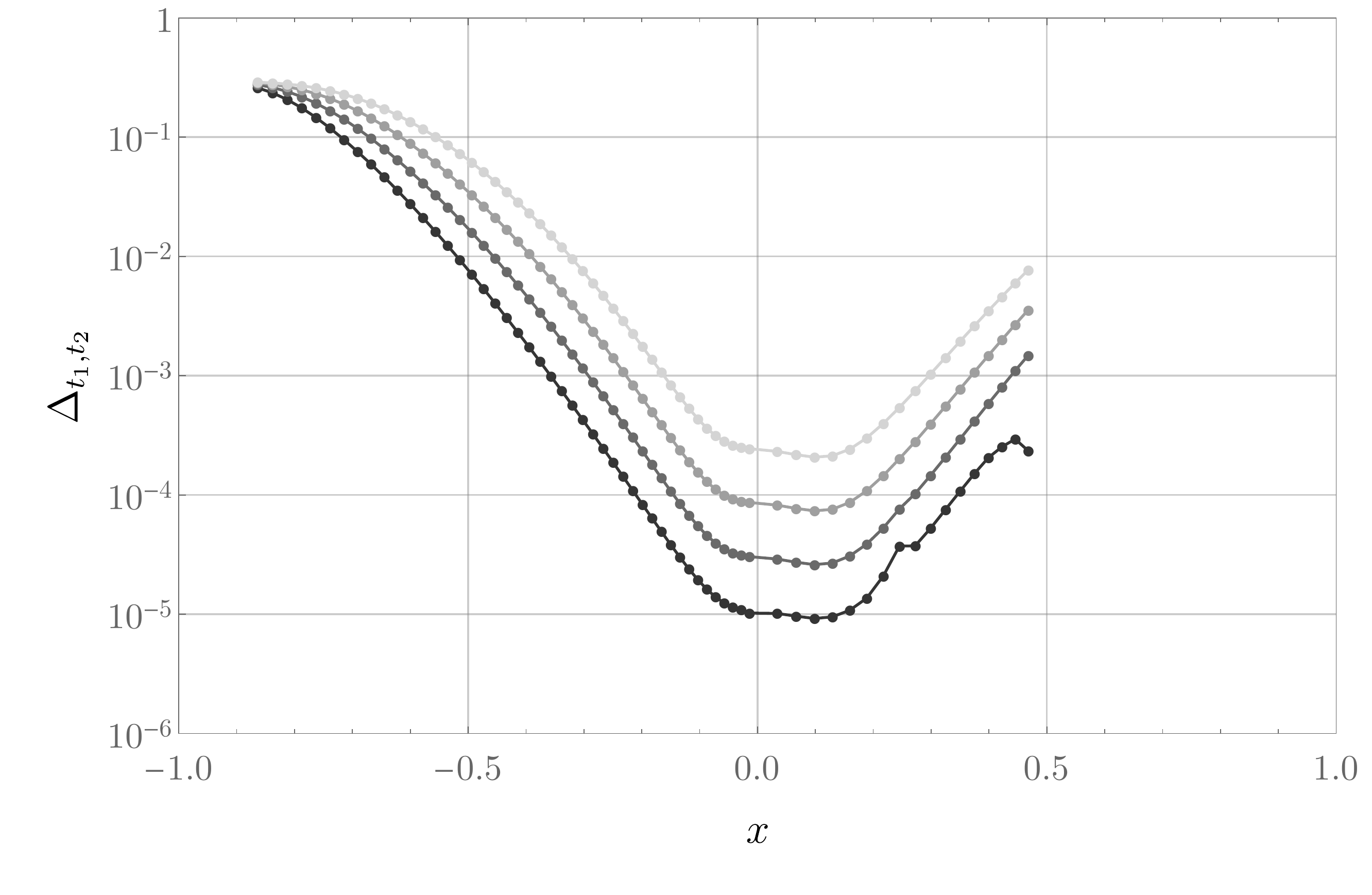}
\label{subfig:tcutoffType1}
}
\subfloat[][Type 2, $n = 1$]{
\includegraphics[width=0.5\textwidth]{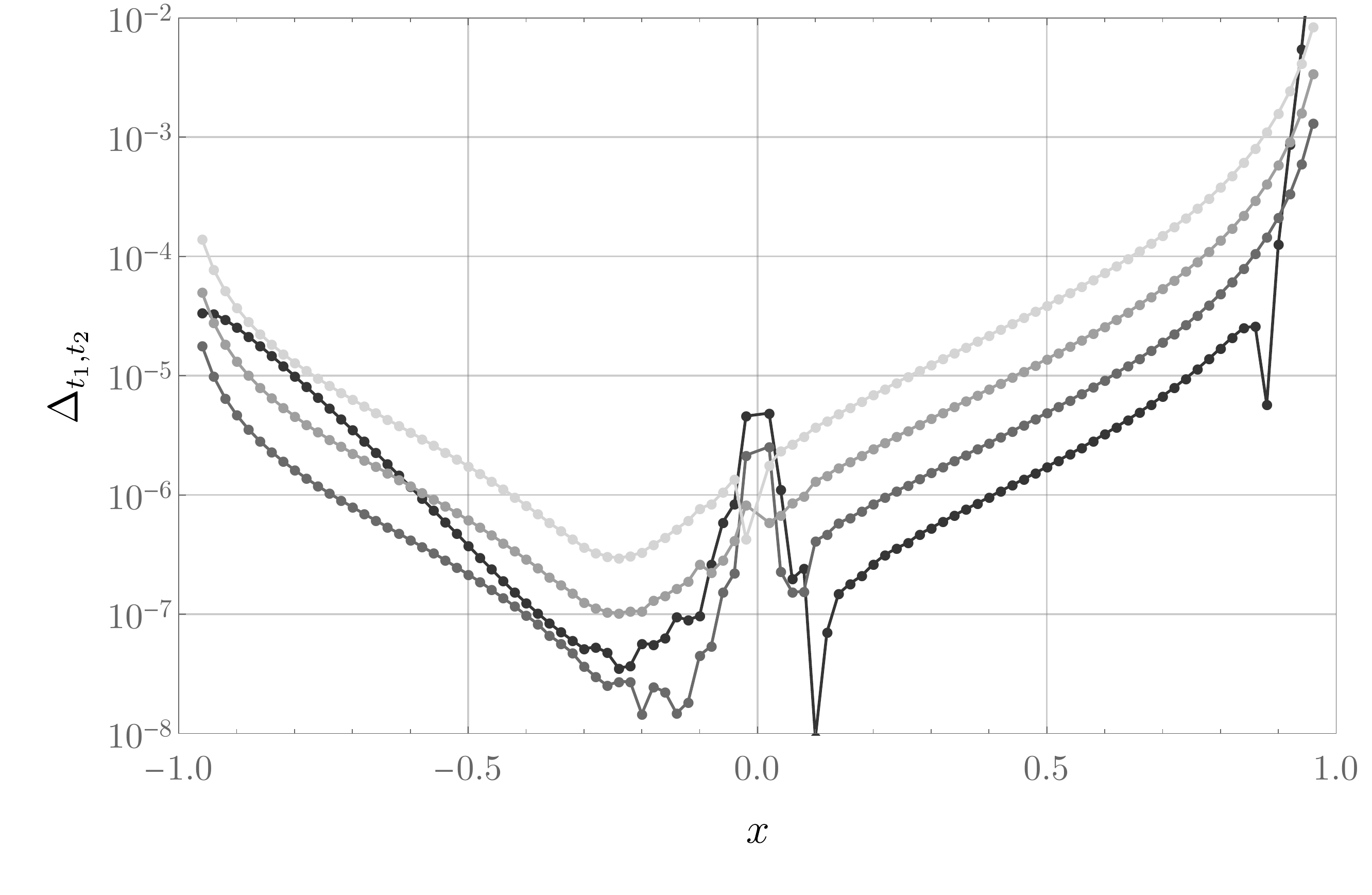}
\label{subfig:tcutoffType2}
}
\caption{Example behavior of~$\Delta_{t_1,t_2}$ for the Dirac fermion (the behavior for the scalar is very similar).  \protect\subref{subfig:tcutoffType1} shows the Type~1 deformation~\eqref{eq:type1metric} with~$l = 6$, while \protect\subref{subfig:tcutoffType2} shows the Type~2 deformation~\eqref{eq:type2metric} with~$n = 1$.  For both we use a grid size of~$N = 800$ and~$t_1$ and~$t_2$ are adjacent elements from the set~$\{0.000125, 0.00025, 0.0005, 0.001, 0.002\}$ with~$t_2 < t_1$; from light gray to black the curves correspond to decreasing~$t_1$,~$t_2$.}
\label{fig:tcutoff}
\end{figure*}

For~$E_{t_*}$ to be a good approximation to the actual value of the Casimir energy, we must include sufficiently many eigenvalues in the traces in~\eqref{eq:truncatedE} that the integrand in the second line recovers the linear behavior~\eqref{eq:smalltK} at~$t \sim t_*$; this is the reason we must include %
hundreds of
thousands of eigenvalues (and therefore 
%
use
a large grid size of~800 points).  For a given geometry, at a fixed and sufficient grid size~$N$, decreasing~$t_*$ should then show convergent behavior before we eventually reach values of~$t_*$ smaller than those accessible at the given resolution, and the convergence should cease.  In this way we may determine the optimal cutoff~$t_*$ for a given~$N$.  To quantify this behavior, we define
\be
\label{eq:Deltat1t2}
\Delta_{t_1,t_2} = \left|1 - \frac{E_{t_* = t_1}}{E_{t_* = t_2}}\right|
\ee
as the relative change between two choices of cutoff.  In Figure~\ref{fig:tcutoff} we show examples of how~$\Delta_{t_1,t_2}$ varies as~$t_1$ and~$t_2$ are decreased for the Dirac fermion at our fiducial grid size of~$N = 800$; Figure~\ref{subfig:tcutoffType1} shows the worst behavior of any of our deformations, while Figure~\ref{subfig:tcutoffType2} shows the best.  The scalar field exhibits analogous behavior, so we do not include additional plots showing it.
It is worth noting that while the holographic gravity solutions generally limit the range of geometries we can access for all the theories, even in the free cases systematic error can become large for the more singular geometries as illustrated in~\ref{subfig:tcutoffType1} for $x \sim -1$. 

\begin{figure*}[t]
\centering
\subfloat[][Type 1, $l = 6$]{
\includegraphics[width=0.5\textwidth]{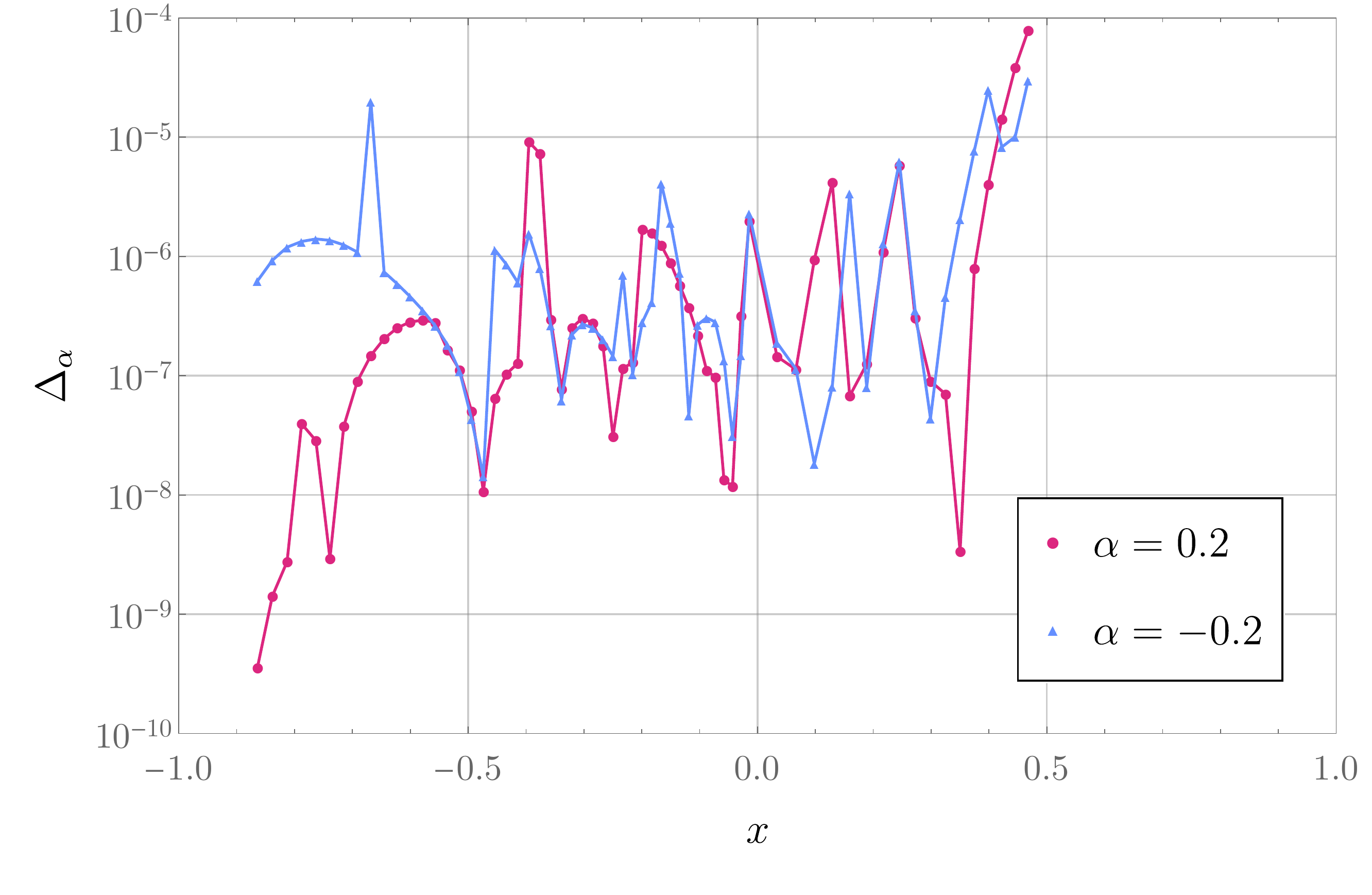}
\label{subfig:DiracalphaType1}
}
\subfloat[][Type 2, $n = 1$]{
\includegraphics[width=0.5\textwidth]{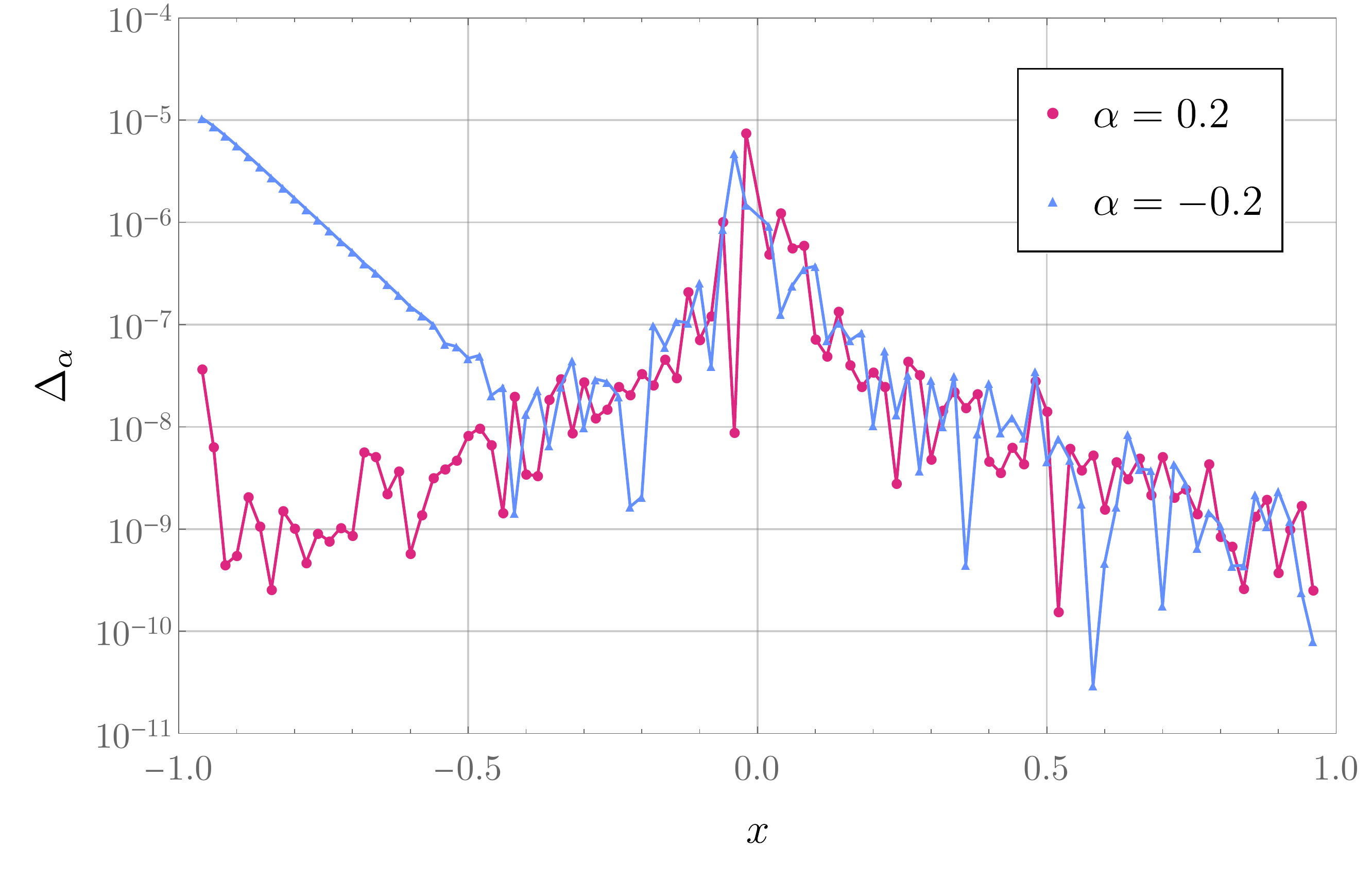}
\label{subfig:DiracalphaType2}
}
\caption{The relative error~$\Delta_\alpha$ of the fermion Casimir energy for~$\alpha = \pm 0.2$ for two representative deformations, computed with a resolution of~$N = 800$.  Figure~\protect\subref{subfig:DiracalphaType1} shows the Type~1 deformation with~$l = 6$, while Figure~\protect\subref{subfig:DiracalphaType2} shows the Type~2 deformation with~$n = 1$.}
\label{fig:Diracalpha}
\end{figure*}

As an additional check of the numerics, we also perform the same computation of the energy under a change of coordinates
\be
\label{eq:coordchange}
\theta' = \theta + \alpha \sin^{3}(2\theta),
\ee
where~$\alpha$ is a parameter we are free to vary.  Of course the energy should be independent of any change of coordinates (and therefore independent of~$\alpha$); any variation with~$\alpha$ therefore offers another estimate of numerical accuracy.  We therefore introduce the relative error
\be
\label{eq:relativeerror}
\Delta_\alpha \equiv \left| 1 - \frac{E_{\alpha}}{E_{\alpha=0}}\right|,
\ee
where~$E_\alpha$ denotes the Casimir energy computed using the deformed coordinate~$\theta'$ defined by~\eqref{eq:coordchange}.  In Figure~\ref{fig:Diracalpha} we show this relative error for the fermion for two representative deformations. While the values of $\alpha = \pm 0.2$ na\"ively seem quite small, if one looks explicitly at the transformed metric function $b(\theta')$, $s(\theta')$ in the new coordinate~$\theta'$, they are very substantially changed from those in the original~$\theta$ coordinate.  The good numerical independence on~$\alpha$ is then excellent confirmation that the results are reliable and accurate.

\subsection{Holographic CFT}
\label{subapp:holo}

\begin{figure*}[t]
\centering
\subfloat[][Type 1, $l = 6$, $x = 0.3$]{
\includegraphics[width=0.5\textwidth]{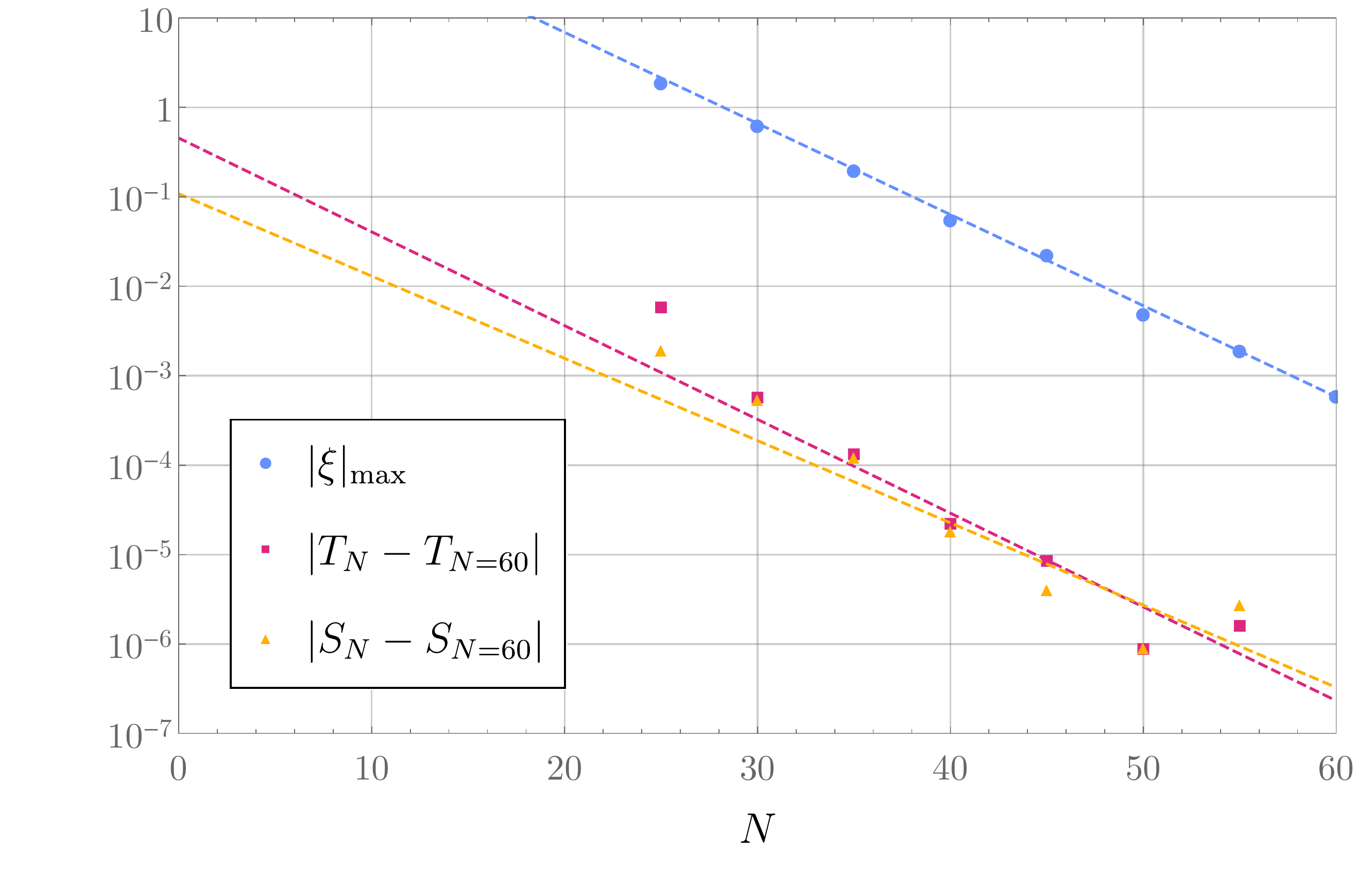}
\label{subfig:PDEType1}
}
\subfloat[][Type 2, $n = 1$, $x = -0.84$]{
\includegraphics[width=0.5\textwidth]{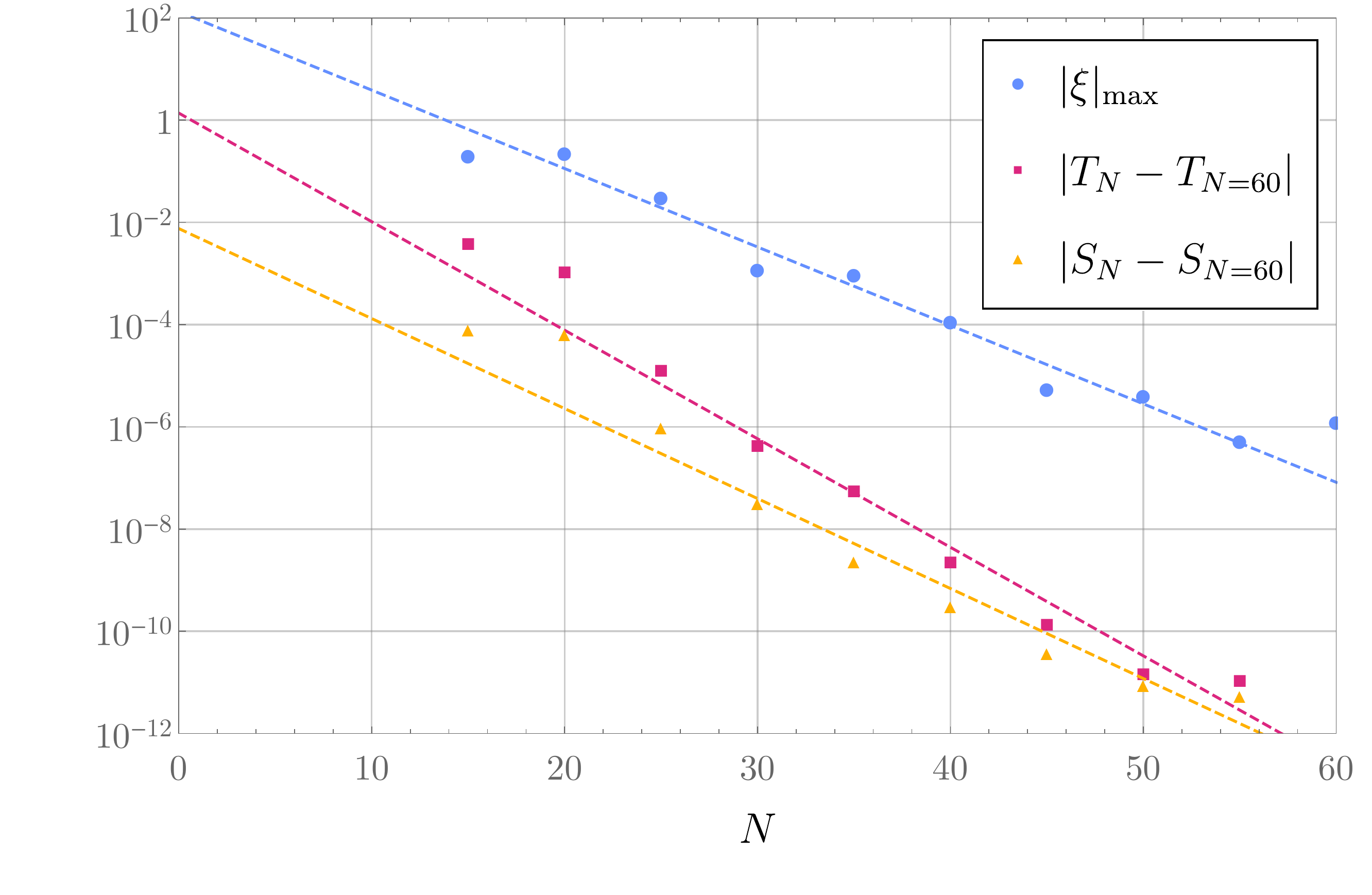}
\label{subfig:PDEType2}
}
\caption{The vanishing of the DeTurck vector~$\xi^a$ (blue circles), as well as the convergence of the metric functions~$T$ and~$S$ (magenta triangles and orange squares) evaluated at the coordinate origin, with increasing numerical resolution.  Figure~\protect\subref{subfig:PDEType1} shows the Type~1 deformation~\eqref{eq:type1metric} with~$l = 6$ and~$x = 0.3$, while Figure~\protect\subref{subfig:PDEType2} shows the Type~2 deformation~\eqref{eq:type2metric} with~$n = 1$ and~$x = -0.84$.  The dashed best-fit lines show the expected exponential convergence.}
\label{fig:PDEconvergence}
\end{figure*}

To assess the accuracy of our numerical calculations of the bulk geometry and of the Casimir energy of the holographic CFT, we perform two checks.  A standard check is to observe the vanishing of the DeTurck vector~$\xi^a$, as well as the convergence of the metric functions~$T$,~$A$,~$B$,~$F$, and~$S$, with increasing grid size~$N$.  To that end, in Figure~\ref{fig:PDEconvergence} we exhibit this convergence for two representative deformations.  This figure shows the maximum value of~$|\xi| \equiv \sqrt{\xi^2}$ on the computational domain, as well as the differences~$T_N(0,0) - T_{N = 60}(0,0)$ and~$S_N(0,0) - S_{N = 60}(0,0)$ between the values of the metric functions~$T$ and~$S$ evaluated at the coordinate origin~$(r,\theta) = (0,0)$ and their values at our highest resolution~$N = 60$.  We also show best-fit lines to the large-$N$ behavior, which are consistent with the exponential convergence expected from pseudospectral methods.

\begin{figure*}[t]
\centering
\subfloat[][Type 1, $l = 6$, $x = 0.3$]{
\includegraphics[width=0.5\textwidth]{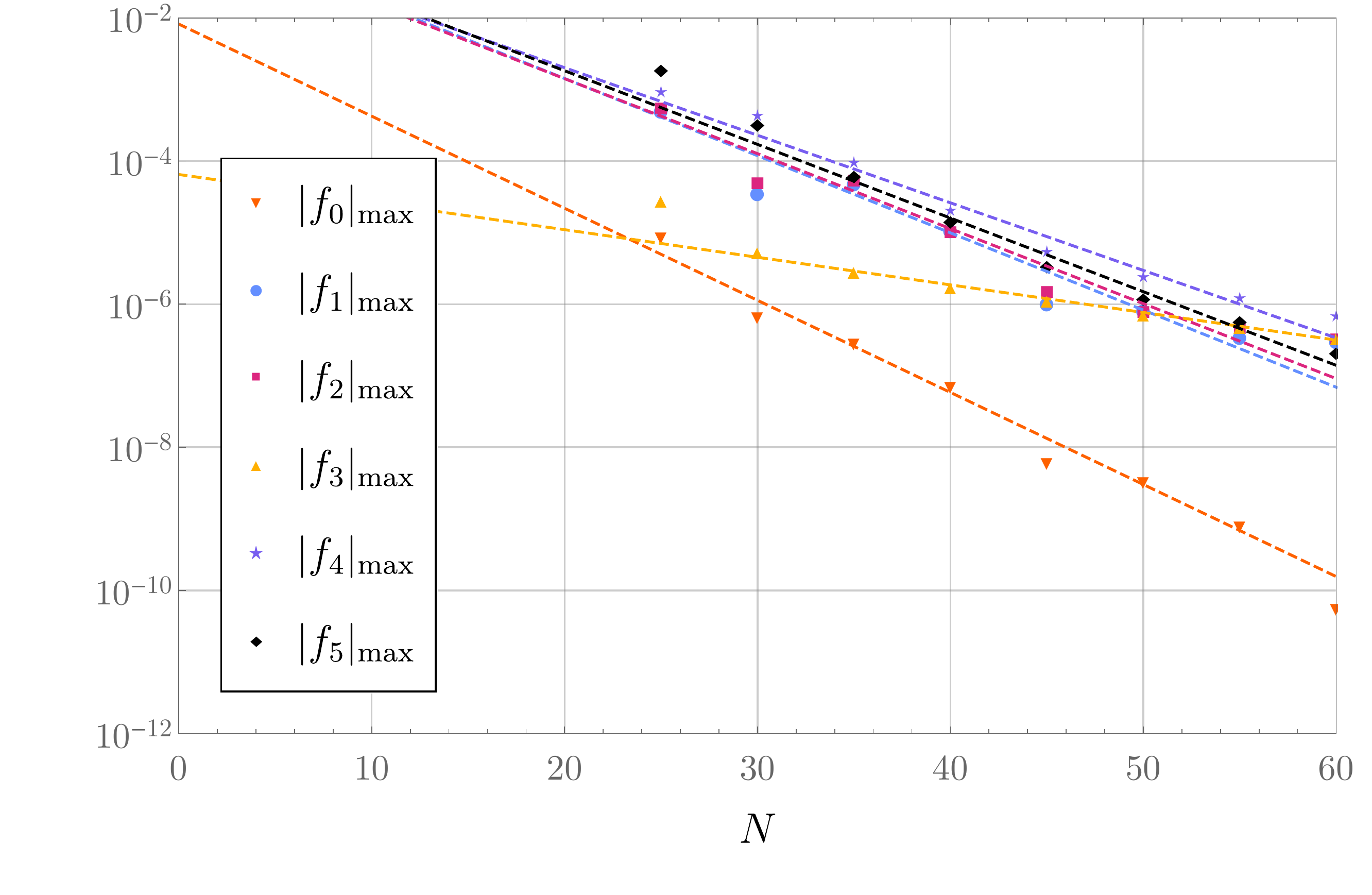}
\label{subfig:fType1}
}
\subfloat[][Type 2, $n = 1$, $x = -0.84$]{
\includegraphics[width=0.5\textwidth]{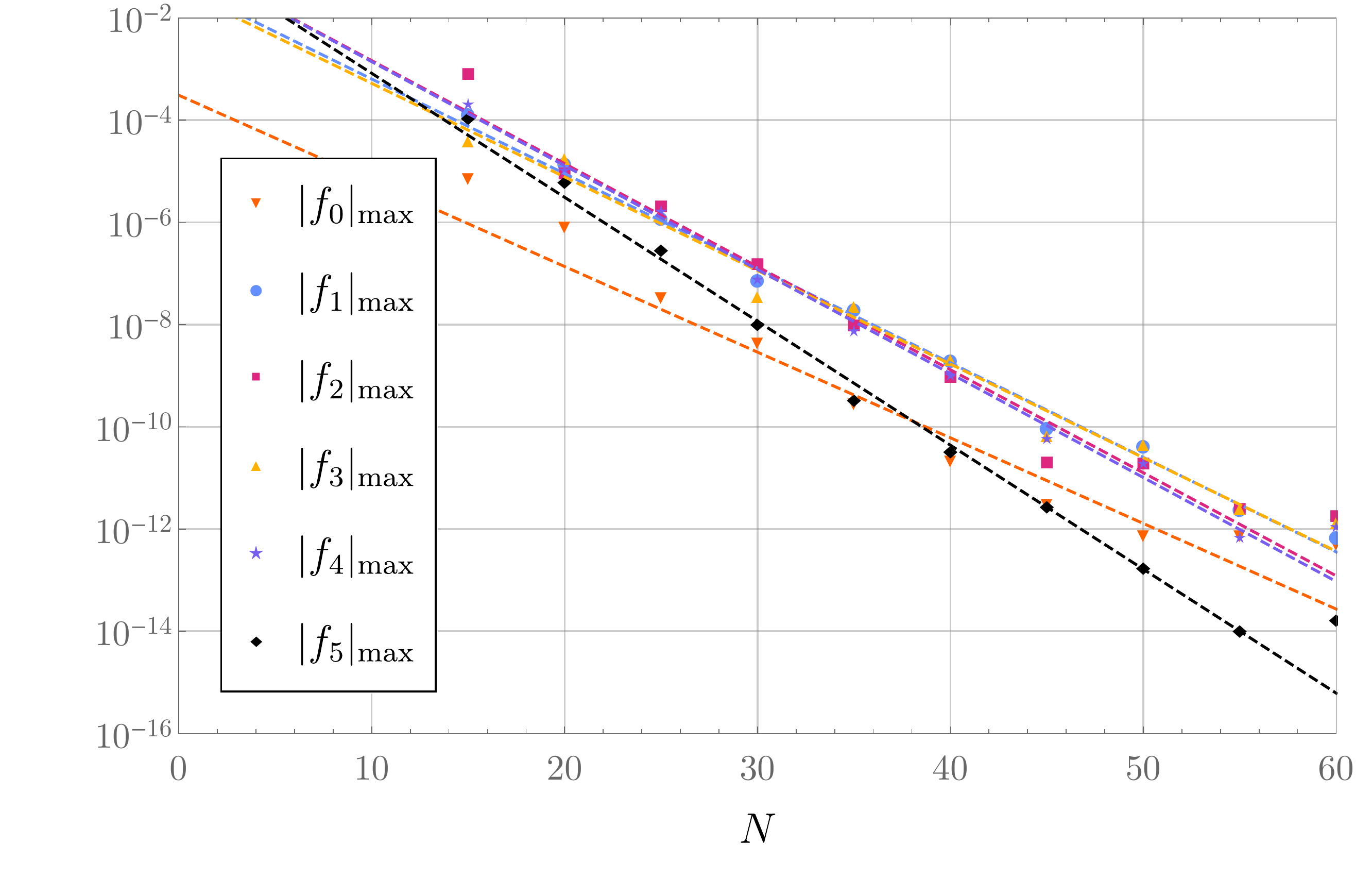}
\label{subfig:fType2}
}
\caption{The maximum absolute values of the functions~$f_0$,~$f_1$,~$f_2$,~$f_3$, and~$f_4$ at the grid points~$r = 0$, and of~$f_5$ at the grid points~$\theta = 0$.  Figure~\protect\subref{subfig:PDEType1} shows the Type~1 deformation~\eqref{eq:type1metric} with~$l = 6$ and~$x = 0.3$, while Figure~\protect\subref{subfig:PDEType2} shows the Type~2 deformation~\eqref{eq:type2metric} with~$n = 1$ and~$x = -0.84$.  Note that~$|f_i|_\mathrm{max}$ vanishes with increasing numerical resolution, indicating convergence to a smooth bulk solution.}
\label{fig:fconvergence}
\end{figure*}

As discussed above, 
%
regularity
 at the origin requires the metric functions to have the expansion~\eqref{eqs:smooth} around~$r = 0$, which requires the vanishing of the functions
\bea
f_0 & = T - T(\theta = 0), \\
f_1 &= A - \frac{1}{2} \left( A_+ +  A_- \cos2\theta\right), \\
f_2 &= B - \frac{1}{2} \left( A_+ -  A_- \cos2\theta\right), \\
f_3 &= S - A(\theta = \pi/2), \\
f_4 &= F +  A_- \cos\theta
\eea
at~$r = 0$, with~$A_\pm \equiv A(\theta = 0) \pm A(\theta = \pi/2)$.  In addition, smoothness at the axis of symmetry~$\theta = 0$ requires
\be
f_5 =  B - S
\ee
to vanish at $\theta = 0$.  In Figure~\ref{fig:fconvergence}, we plot $\max_\mathrm{grid} | f_i |$, the maximum absolute value of the various functions $f_i$ over the grid points $r = 0$ (for $f_0$ - $f_4$) or $\theta = 0$ (for $f_5$) for representative solutions. These show that indeed these functions all vanish exponentially with increasing grid size, compatible with convergence to a smooth solution.

\begin{figure*}[t]
\centering
\subfloat[][Type 1, $l = 6$, $x = 0.3$]{
\includegraphics[width=0.5\textwidth]{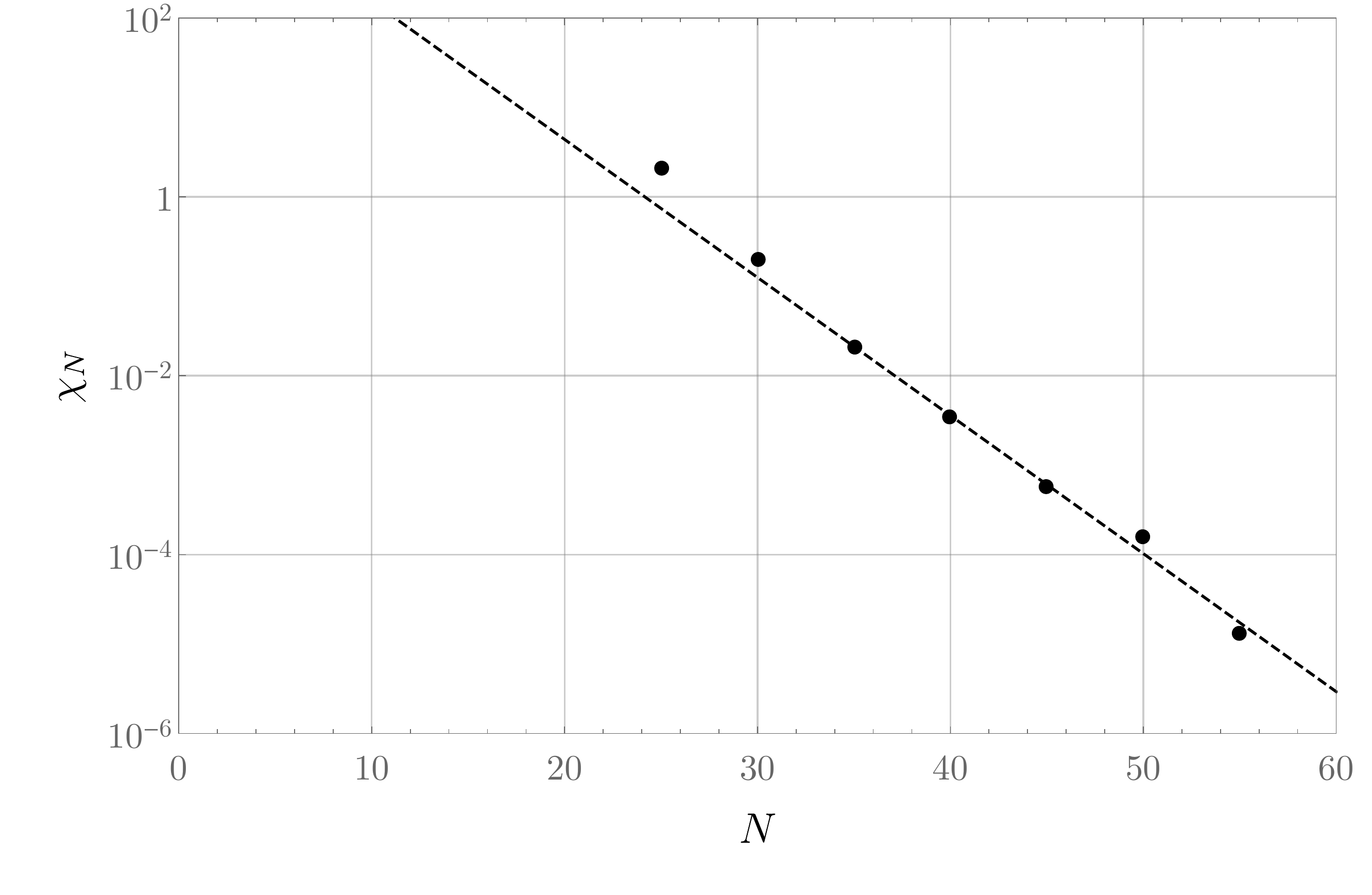}
\label{subfig:EnergyType1}
}
\subfloat[][Type 2, $n = 1$, $x = -0.84$]{
\includegraphics[width=0.5\textwidth]{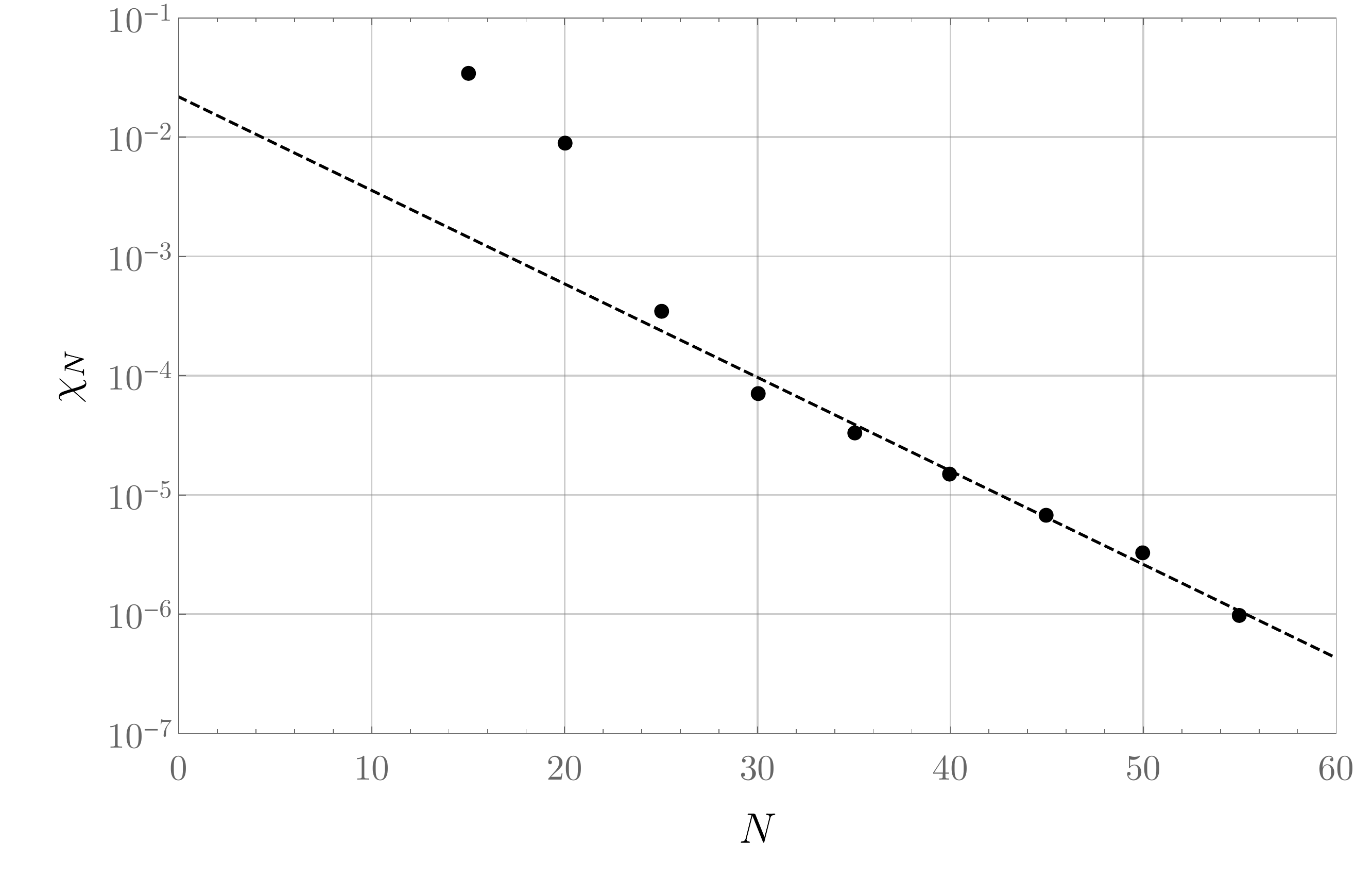}
\label{subfig:EnergyType2}
}
\caption{The vanishing of the relative error~$\chi_N$ in the Casimir energy with increasing numerical resolution.  Figure~\protect\subref{subfig:EnergyType1} shows the Type~1 deformation~\eqref{eq:type1metric} with~$l = 6$ and~$x = 0.3$, while Figure~\protect\subref{subfig:EnergyType2} shows the Type~2 deformation~\eqref{eq:type2metric} with~$n = 1$ and~$x = -0.84$.  The dashed best-fit lines show the expected exponential convergence.}
\label{fig:Energyconvergence}
\end{figure*}

Since we are ultimately interested in the Casimir energy computed from these gravitational solutions (obtained using~\eqref{eq:Egrav}), we also compute the relative error
\be
\chi_N \equiv \left|1 - \frac{E_N}{E_{N = 60}}\right|
\ee
between the energy~$E_N$ with grid size~$N$ and that obtained with the maximum grid size of~$N = 60$.  This relative error is plotted in Figure~\ref{fig:Energyconvergence} for the same two representative deformations, and also exhibits the expected exponential convergence in~$E_N$.

\begin{figure*}[t]
\centering
\subfloat[][Type 1, $l = 6$]{
\includegraphics[width=0.5\textwidth]{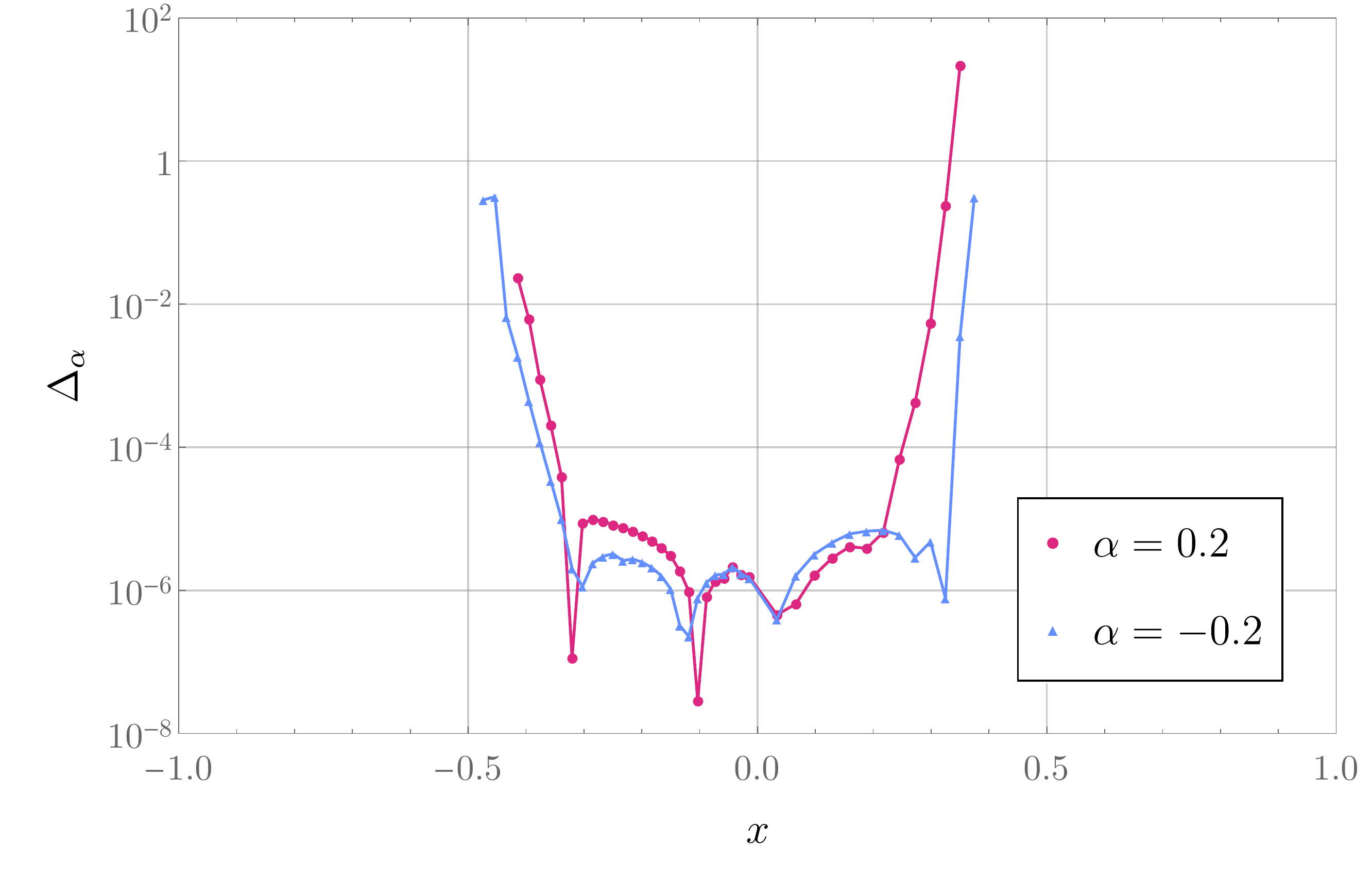}
\label{subfig:gravalphaworst}
}
\subfloat[][Type 2, $n = 1$]{
\includegraphics[width=0.5\textwidth]{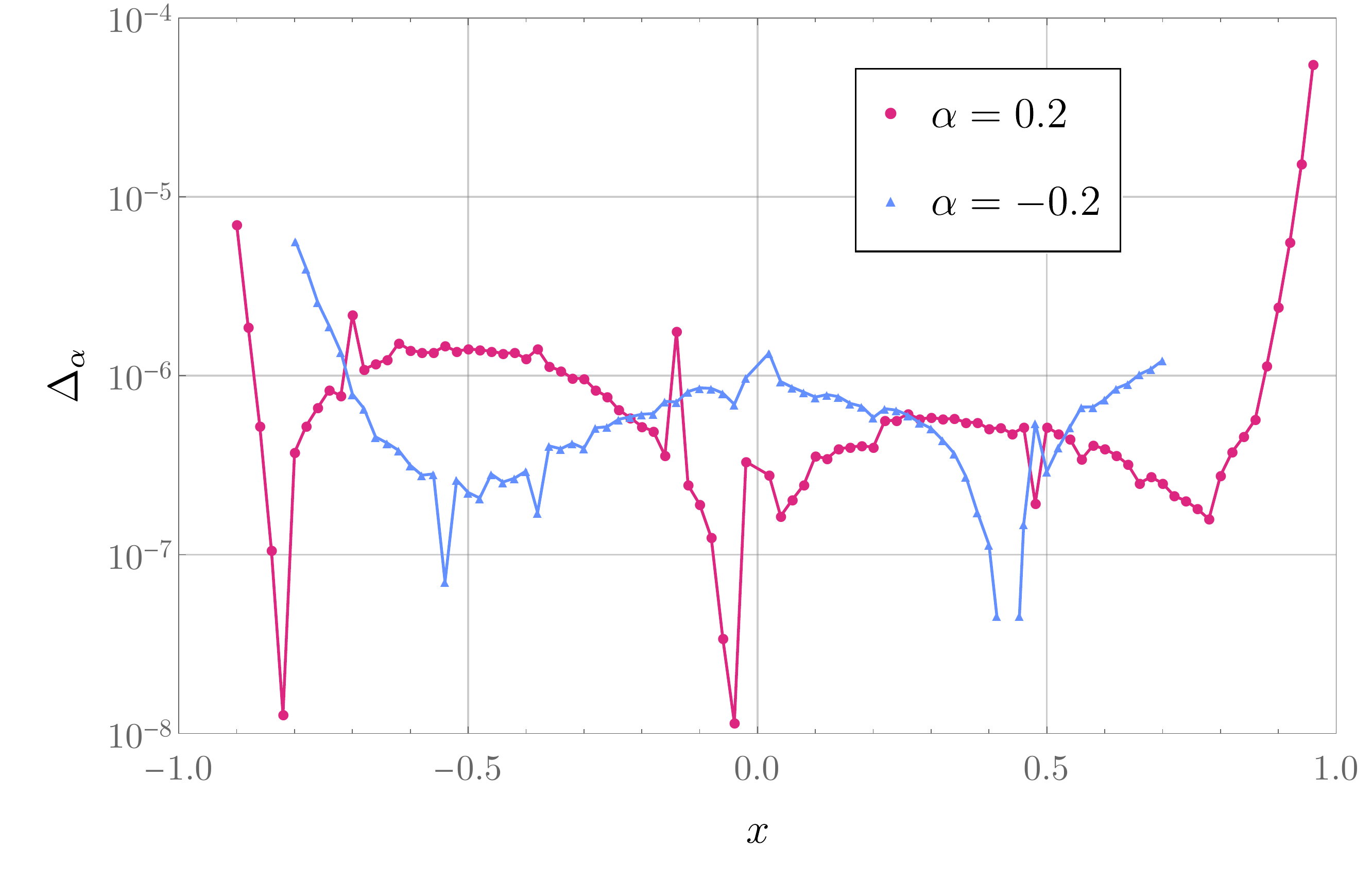}
\label{subfig:gravalphabest}
}
\caption{The relative error~$\Delta_\alpha$ for~$\alpha = \pm 0.2$ for two representative deformations, computed with a resolution of~$N = 60$.  Figure~\protect\subref{subfig:gravalphaworst} shows the Type~1 deformation with~$l = 6$, while Figure~\protect\subref{subfig:gravalphabest} shows the Type~2 deformation with~$n = 1$; these correspond to the worst- and best-behaved deformations, respectively.}
\label{fig:gravalpha}
\end{figure*}

As an additional check of the numerics, we also perform the same computation of the energy under the change of coordinates~\eqref{eq:coordchange} described above.  In Figure~\ref{fig:gravalpha} we show representative plots of the relative error~$\Delta_\alpha$ (defined in~\eqref{eq:relativeerror}) for the gravitational solutions, highlighting the best and worst behaviors exhibited by any of our deformations.
%
We have also computed the same energies using different choices of the reference metric interpolation function $P(r)$, and again the results are independent of this as they should be.

%
\bibliographystyle{jhep}
\bibliography{deformedAdS}
%

\end{document}